\documentclass[iop,apj]{emulateapj}
\usepackage{apjfonts}
\usepackage{graphicx}
\usepackage{amsmath}
\usepackage{amsbsy}
\newcommand{\hetwolong}{\mbox{HE\,$0230-2130$}}
\newcommand{\mgfourlong}{\mbox{MG\,J$0414+0534$}}
\newcommand{\hefourlong}{\mbox{HE\,$0435-1223$}}
\newcommand{\rxninelong}{\mbox{RX\,J$0911+0551$}}
\newcommand{\sdssninelong}{\mbox{SDSS\,J$0924+0219$}}
\newcommand{\heoneonelong}{\mbox{HE\,$1113-0641$}}
\newcommand{\pgoneonelong}{\mbox{PG\,$1115+080$}}
\newcommand{\rxoneonelong}{\mbox{RX\,J$1131-1231$}}
\newcommand{\sdssoneonelong}{\mbox{SDSS\,J$1138+0314$}}
\newcommand{\sdssonethreelong}{\mbox{SDSS\,J$1330+1810$}}
\newcommand{\wfitwosixlong}{\mbox{WFI\,J$2026-4536$}}
\newcommand{\wfithreelong}{\mbox{WFI\,J$2033-4723$}}
\newcommand{\hetwo}{\mbox{HE\,$0230$}}
\newcommand{\mgfour}{\mbox{MG\,J$0414$}}
\newcommand{\hefour}{\mbox{HE\,$0435$}}
\newcommand{\rxnine}{\mbox{RX\,J$0911$}}
\newcommand{\sdssnine}{\mbox{SDSS\,J$0924$}}
\newcommand{\heoneone}{\mbox{HE\,$1113$}}
\newcommand{\pgoneone}{\mbox{PG\,$1115$}}
\newcommand{\rxoneone}{\mbox{RX\,J$1131$}}
\newcommand{\sdssoneone}{\mbox{SDSS\,J$1138$}}
\newcommand{\sdssonethree}{\mbox{SDSS\,J$1330$}}
\newcommand{\wfitwosix}{\mbox{WFI\,J$2026$}}
\newcommand{\wfithree}{\mbox{WFI\,J$2033$}}

\newcommand{\qtwotwo}{\mbox{Q\,$2237+0305$}}

\newcommand{\chandra}{\textit{Chandra}}
\newcommand{\cxo}{\textit{Chandra X-ray Observatory}}

\newcommand{\ergcms}{\ensuremath{\mathrm{erg~cm}^{-2}~\mathrm{s}^{-1}}}

\newcommand{\err}[2]{\ensuremath{^{_{+#1}}_{^{-#2}}}}
\newcommand{\tnm}[1]{\tablenotemark{#1}}

\begin{document}
\bibliographystyle{apj}
\shorttitle{SIZES AND TEMPERATURE PROFILES OF QUASAR ACCRETION DISKS}
\shortauthors{BLACKBURNE ET AL.}
\slugcomment{}
\title{Sizes and Temperature Profiles of Quasar Accretion Disks \\
from Chromatic Microlensing$^1$}

\author{Jeffrey~A.~Blackburne\altaffilmark{2,3},
  David~Pooley\altaffilmark{4}, Saul~Rappaport\altaffilmark{3},
  \& Paul~L.~Schechter\altaffilmark{3}}

\altaffiltext{1}{This paper includes data gathered with the 6.5 m
  Magellan Telescopes located at Las Campanas, Chile.}
\altaffiltext{2}{Department of Astronomy and Center for Cosmology and
  AstroParticle Physics, The Ohio State University, 140 West 18th
  Avenue, Columbus, OH 43210, USA; blackburne@astronomy.ohio-state.edu}
\altaffiltext{3}{Massachusetts Institute of Technology, Department of
  Physics and Kavli Institute for Astrophysics and Space Research, 70
  Vassar Street, Cambridge, MA 02139, USA}
\altaffiltext{4}{Eureka Scientific, 2452 Delmer Street Suite 100, Oakland,
  CA 94602, USA}
\begin{abstract}

Microlensing perturbations to the flux ratios of gravitationally
lensed quasar images can vary with wavelength because of the chromatic
dependence of the accretion disk's apparent size. Multiwavelength
observations of microlensed quasars can thus constrain the temperature
profiles of their accretion disks, a fundamental test of an important
astrophysical process which is not currently possible using any other
method. We present single-epoch broadband flux ratios for 12 quadruply
lensed quasars in 8 bands ranging from 0.36 to 2.2 $\mu$m, as well as
{\it Chandra} \mbox{0.5 -- 8\,keV} flux ratios for five of them. We
combine the optical/IR and X-ray ratios, together with X-ray ratios
from the literature, using a Bayesian approach to constrain the
half-light radii of the quasars in each filter. Comparing the overall
disk sizes and wavelength slopes to those predicted by the standard
thin accretion disk model, we find that on average the disks are
larger than predicted by nearly an order of magnitude, with sizes that
grow with wavelength with an average slope of $\sim$0.2 rather than
the slope of $4/3$ predicted by the standard thin disk theory. Though
the error bars on the slope are large for individual quasars, the
large sample size lends weight to the overall result. Our results
present severe difficulties for a standard thin accretion disk as the
main source of UV/optical radiation from quasars.

\end{abstract}


\section{Introduction}
\label{sec:intro}

Microlensing of the images of gravitationally lensed quasars by stars
in the foreground galaxies causes time-variable anomalies in their
flux ratios \citep[see review by][]{Wambsganss:2006p453}. In-depth
study of these anomalies offers a unique glimpse into the properties
of both the deflector and the quasar itself. Because the amplitude of
microlensing anomalies depends on the angular size of the source, it
can be used to estimate the size of quasar emission regions at the
level of microarcseconds, far beyond current resolution limits
\citep[e.g.,][]{Rauch:1991pL39}. The temperature structure of quasar
accretion disks on these same scales causes chromatic microlensing,
where blue light from the inner regions is more strongly microlensed
than red light from farther out \citep{Wambsganss:1991p864}. Thus,
multiwavelength observations of quasar microlensing have a unique
capability to measure the structure of accretion disks and put direct
constraints on accretion models.

Several recent studies have made use of microlensing to explore the
size scale of the accretion disk, which emits most of the optical
light. \citet{Pooley:2007p19} follow up on earlier results
\citep{Blackburne:2006p569, Pooley:2006p67} to show that the flux
ratio anomalies caused by microlensing are in general stronger in
X-rays than at optical wavelengths, and infer that the source of
optical light is larger than the X-ray source, and larger on average
than expected for a standard thin accretion disk model
\citep{Shakura:1973p337, Novikov:1973p343}. \citet{Morgan:2010p1129}
estimate disk sizes for several lensed quasars using time-series
observations of microlensing variability, rather than single-epoch
flux ratio anomalies. They find only marginally better agreement with
thin disk models, and note that flux-based size predictions are
systematically too small.

Several other studies move beyond single-wavelength size estimates to
estimate the dependence of the source size on wavelength and thus
constrain the temperature profile of the disk. \citet{Bate:2008p1955}
and \citet{Floyd:2009p233} place upper limits on the radii of
accretion disks using single-epoch broadband flux ratio measurements;
they also put weak constraints on the temperature profile
slope. \citet{Mosquera:2010p0} make size and slope estimates using
narrowband photometry. Other studies combine multiwavelength
observations with time-series light curves \citep{Anguita:2008p327,
Poindexter:2008p34, Mosquera:2009p1292, Poindexter:2010p668}, or even
employ spectroscopic monitoring \citep{Eigenbrod:2008p933}.
Additionally, efforts have been made \citep{Morgan:2008p755,
Dai:2010p278} to estimate the size of the source of the non-thermal
X-ray flux by combining \chandra\ measurements with optical light
curves. Because of the effort involved in these observations, each of
these studies focuses on an individual lensed quasar.

The effects of microlensing also depend on properties of the
foreground galaxy---the mass function of the stars, and the surface
density of clumpy stellar matter as a fraction of the total (smooth
and clumpy) surface density. The amplitude of microlensing
perturbations decreases for very small stellar fractions, and also
decreases, perhaps counterintuitively, for large stellar fractions
\citep{Schechter:2002p685}. This has led to results that effectively
rule out the 100\% stars case as well as the 100\% smooth matter case,
with a best-fit stellar mass fraction of $\sim$10\%
\citep{Schechter:2004p103, Pooley:2009p1892}. The mass function of the
microlens stars changes the variability characteristics of
microlensing light curves; \citet{Poindexter:2010p658} exploit this to
estimate the mean stellar mass in the lensing galaxy of the quadruple
lens \qtwotwo.

With the goal of measuring accretion disk sizes and temperature
profiles, we present single-epoch observations of the flux ratios of a
sample of 12 lensed quasars in eight broadband optical and infrared (IR)
filters in the ``big blue bump'' region of the spectrum. The low cost
of our single-epoch observing strategy relative to long monitoring
campaigns allows us to cover a factor of six in wavelength, while
roughly tripling the number of lensed quasars with observations of
this kind. We also present X-ray flux ratios from the \cxo\ for 5 of
the quasars. Comparing these flux ratios, as well as X-ray ratios from
the literature for six of the seven remaining lenses, to those predicted by
smooth lens models, we use a Bayesian analysis method to determine
posterior probability distributions for the sizes of the quasar
emission regions in each filter. This allows us to estimate the
overall size scales of the accretion disks, as well as the scaling of
the size with black hole mass and wavelength. As we will see, the
relatively large size of our sample is crucial for drawing conclusions
about accretion disk structure.

Our sample is entirely made up of quadruply lensed quasars; they have
the advantage of having three flux ratios apiece (rather than one for
double lenses) and are easier to model. They are as follows:\\
\hetwolong\ \citep{Wisotzki:1999pL41},\\
\mgfourlong\ \citep{Hewitt:1992p968},\\
\hefourlong\ \citep{Wisotzki:2002p17},\\
\rxninelong\ \citep{Bade:1997pL13},\\
\sdssninelong\ \citep{Inada:2003p666},\\
\heoneonelong\ \citep{Blackburne:2008p374},\\
\pgoneonelong\ \citep{Weymann:1980p641},\\
\rxoneonelong\ \citep{Sluse:2003pL43},\\
\sdssoneonelong\ \citep{Eigenbrod:2006p759},\\
\sdssonethreelong\ \citep{Oguri:2008p1973},\\
\wfitwosixlong\ \citep{Morgan:2004p2617}, and\\ 
\wfithreelong\ \citep{Morgan:2004p2617}.\\
For brevity, we will usually exclude the declination from their
names from now on. Following \citet{Pooley:2007p19}, we categorize the
quasar images by their macro-magnification and parity: highly
magnified minima and saddle points of the light travel time surface
are designated HM and HS, while their less-magnified counterparts are
LM and LS, respectively.

The variation of accretion disk size with black hole mass and
wavelength provides direct constraints on disk models. The standard
thin disk model predicts a profile for the effective temperature that
falls approximately as the $\beta = 3/4$ power of the radius:
\begin{equation}
\label{eqn:temp}
T_\mathrm{eff}(r) = \left(\frac{3 G^2 M_\mathrm{BH}^2 m_p f_\mathrm{Edd}}
{2 c \sigma_B \sigma_T \eta r^3}\right)^{1/4} g(r_\mathrm{in}/r)^{1/4} ~.
\end{equation}
where $M_\mathrm{BH}$ is the black hole mass, $m_p$ is the proton
mass, $\sigma_B$ is the Stefan-Boltzmann constant, and $\sigma_T$ is
the Thomson scattering cross-section. The dimensionless factors
$f_\mathrm{Edd}$ and $\eta$ are the ratios of the quasar's bolometric
luminosity to its Eddington luminosity and to the accretion rate
$\dot{M}c^2$, respectively. The function $g$ is defined as $g(x) =
1-x^{1/2}$, and $r_\mathrm{in}$ is the innermost radius of the
accretion disk. As we will see, at optical wavelengths the disk size
is much larger than $r_\mathrm{in}$, so for simplicity we set $g$ to
unity. Likewise, the general relativistic corrections of
\citet{Novikov:1973p343} are small at the radii of interest. By
integrating the specific flux at fixed wavelength over the entire
multi-temperature disk, we calculate the half-light radius---that
is, the radius at which half of the light at a given wavelength is
emitted---predicted by the thin disk temperature profile:
\begin{align}
\label{eqn:rhalf}
r_{1/2} &= 2.44 \notag
\left[\frac{45 G^2 M_\mathrm{BH}^2 m_p f_\mathrm{Edd} \lambda^4}
{4 \pi^5 h_P c^3 \sigma_T \eta}\right]^{1/3} \sqrt{\cos{i}} \\
&= 1.68 \times 10^{16} \mathrm{cm}
\left(\frac{M_\mathrm{BH}}{10^9 M_\odot}\right)^{2/3}
\left(\frac{f_\mathrm{Edd}}{\eta}\right)^{1/3}
\left(\frac{\lambda}{\mu\mathrm{m}}\right)^{4/3} ~.
\end{align}
In this equation, $h_P$ is Planck's constant, and $i$ is the
unknown inclination angle. We include the factor of $(\cos i)^{1/2}$
to account for the inclination of the disk to our line of sight, and
set it to a reasonable average value of $(1/2)^{1/2}$ in the second
line. For a thin disk, the half-light radius scales with the rest
wavelength as $\lambda^{4/3}$; for a more general disk temperature
profile $T_\mathrm{eff} \propto r^{-\beta}$, the half-light radius
will depend on $\lambda^{1/\beta}$.

In Sections~\ref{sec:xray-obs} and \ref{sec:photometry}, we describe
the X-ray and optical/IR observations and flux ratio
measurements. In Section~\ref{sec:uncertainty}, we estimate the
systematic uncertainties in the measured flux ratios. We describe our
lens models in Section~\ref{sec:models}, and in
Section~\ref{sec:analysis} we describe our Bayesian analysis
method. We discuss our results and give concluding remarks in
Sections~\ref{sec:results} and \ref{sec:conclusions}. Throughout this
work, we calculate distances and time delays using a geometrically
flat universe with $\Omega_M = 0.3$, $\Omega_\Lambda = 0.7$, and $H_0
= 70$ km s$^{-1}$ Mpc$^{-1}$.


\section{X-ray Flux Ratios}
\label{sec:xray-obs}

\begin{deluxetable*}{rlrclllcc}
\tablewidth{0pt}
\tablecaption{X-ray Fluxes and Flux Ratios\label{tab:xraydata}}
\tablehead{
  \multicolumn{3}{l}{Quasar} & 
  & 
  \multicolumn{3}{c}{Image Flux Ratios\tnm{a}} & 
  & 
  \colhead{LM Unabs.\ $F_\mathrm{0.5-8\,keV}$\tnm{b}}\\ 
  \cline{1-3} \cline{5-7}\\[-2.5ex]
  \colhead{ObsID\tnm{c}} & 
  \colhead{Date} & 
  \colhead{Exposure\ (s)} & 
  & 
  \colhead{HS/LM} & 
  \colhead{HM/LM} & 
  \colhead{LS/LM} &
  & 
  \colhead{($10^{-14}~\ergcms$)}
}
\startdata   
\multicolumn{3}{l}{\hefourlong} && B/A                     & C/A                     & D/A                     && $F_A$                 \\ \cline{1-3}
7761 & 2006 Dec 17 & 10\,130    && 0.375\err{0.049}{0.045} & 0.378\err{0.049}{0.044} & 0.363\err{0.048}{0.043} && 17.75\err{4.43}{4.40} \\[2ex]
\multicolumn{3}{l}{\heoneonelong} && D/A                  & B/A                  & C/A                  && $F_A$             \\ \cline{1-3}
7760 & 2007 Jan 28 & 15\,180      && 0.78\err{0.32}{0.32} & 0.63\err{0.41}{0.41} & 0.20\err{0.16}{0.16} && 7.8\err{2.3}{2.3} \\[2ex] 
\multicolumn{3}{l}{\rxoneonelong\tnm{d}} && A/C                     & B/C                     & D/C                     && $F_C$                 \\ \cline{1-3}
7787 & 2007 Feb 13 &  5\,190             && 5.717\err{0.417}{0.382} & 3.176\err{0.244}{0.224} & 0.624\err{0.064}{0.058} && 45.34\err{3.99}{3.88} \\
7789 & 2007 Apr 16 &  5\,190             && 5.580\err{0.391}{0.359} & 2.922\err{0.218}{0.200} & 0.374\err{0.044}{0.040} && 47.44\err{4.14}{4.03} \\[2ex]
\multicolumn{3}{l}{\sdssoneonelong} && D/C               & A/C               & B/C               && $F_C$             \\ \cline{1-3}
7759 & 2007 Feb 13 & 19\,080        && 1.3\err{0.6}{0.5} & 3.2\err{1.0}{1.0} & 1.0\err{0.4}{0.4} && 1.6\err{0.9}{0.9} \\[2ex] 
\multicolumn{3}{l}{\wfitwosixlong} && A2/B              & A1/B              & C/B                  && $F_B$             \\ \cline{1-3}
7758 & 2007 Jun 28 & 10\,170       && 2.0\err{1.7}{1.0} & 5.8\err{1.8}{1.7} & 0.40\err{0.06}{0.15} && 5.3\err{1.2}{1.2}
\enddata
\tablenotetext{a}{HS: highly magnified saddle point; HM: highly
  magnified minimum; LS: less magnified saddle point; LM: less
  magnified minimum.}
\tablenotetext{b}{The unabsorbed flux of the LM image is computed from
  the best fit power-law model described in Section
  \ref{sec:xray-obs}.}
\tablenotetext{c}{The observation identifier of the
  \chandra\ dataset.}
\tablenotetext{d}{The data for \rxoneonelong\ come from the \chandra\
  archive.}
\tablecomments{We only report new measurements here; see
  references in the text for other flux ratios.}
\end{deluxetable*}

The X-ray data in this survey are from the Advanced CCD Imaging
Spectrometer (ACIS) instrument aboard the \cxo. For five of the lenses
among our sample (specifically \hetwo, \mgfour, \rxnine, \sdssnine,
and \wfithree) we adopt the X-ray flux ratios reported by
\citet{Pooley:2007p19}. For \pgoneone\ we adopt the X-ray flux ratios
of the 2008 January 31 observation (ObsID 7757) from
\citet{Pooley:2009p1892}, as they are more contemporaneous with our
optical/IR observations. For the same reason, for \rxoneone\ we
use observations from the \chandra\ archive.  Specifically, we use the
2007 February 13 observation (ObsID 7787) for comparison with optical
data and the 2007 April 16 observation (ObsID 7789) for comparison
with IR data (see Section~\ref{sec:photometry}). The high level
of X-ray microlensing variability in these two systems in particular
\citep{Chartas:2009p174, Pooley:2009p1892} underscores the importance
of making our multiwavelength observations as contemporaneous as
possible (see Section~\ref{sec:xrayuncertainty} for a detailed
discussion).

The remaining lenses in our sample, excluding \sdssonethree, which has
not been observed in X-rays, were observed using ACIS between
2006~December~17 and 2007~June~28. The data were reduced using the
procedure described by \citet{Pooley:2009p1892}. We produced
$0.5$--$8$\,keV images of each lens with a resolution of 0\farcs0246
pixel$^{-1}$, and fit four Gaussian components and a constant
background to each image. The background level was fixed at a level
determined from a nearby source-free region, and the relative
positions of the four Gaussians were fixed to the \textit{Hubble Space
  Telescope} (\textit{HST}) positions provided by the CASTLES
database\footnote{See \url{http://www.cfa.harvard.edu/castles}}
\citep{Falco:2001p25}. The width of the Gaussian point-spread function
(PSF) was allowed to vary for each lens, but was constrained to be the
same for all four quasar components. The fits were performed using
\citet{Cash:1979p939} statistics, which are appropriate when small
numbers of photons have been collected. In the cases of \heoneone,
\sdssoneone, and \wfitwosix, the least-squares optimization settled on
a wider PSF than for any of the other observations in our previous
work. Given the low numbers of total counts in these observations and
the small separations of the images, we fixed the width of the PSF in
these cases to the average of the other X-ray PSFs. The flux in image
A2 of \wfitwosix\ was very sensitive to the PSF, so we allocated a
generous 1.5 mag of uncertainty to its flux ratio, added in quadrature
to the other sources of uncertainty (see Section
\ref{sec:xrayuncertainty}). The other flux ratios were not sensitive
to the PSF size. Table \ref{tab:xraydata} summarizes the
\chandra\ observations and the measured flux ratios.


\section{Optical and Infrared Photometry}
\label{sec:photometry}

\begin{deluxetable*}{llcccc}
\tablewidth{0pt}
\tablecaption{Optical Observations
  \label{tab:optobs}}
\tablehead{
  \colhead{Quasar} & 
  \colhead{Date} & 
  \colhead{Instrument} &
  \colhead{Filters} & 
  \colhead{Exposures (s)} & 
  \colhead{Seeing} 
}
\startdata
\hetwo        & 2007 Sep 16 & IMACS & $u'g'r'i'z'$ & $720; 240; 240; 240; 480$  & 0\farcs43 \\
              & 2007 Jul 29 & PANIC & $JHK_s$      & $540; 540; 540$            & 0\farcs64 \\
\mgfour       & 2007 Sep 21 & IMACS & $r'i'z'$     & $360; 240; 480$            & 0\farcs67 \\
              & 2007 Sep 22 & PANIC & $JHK_s$      & $810; 720; 720$            & 0\farcs76 \\
\hefour       & 2007 Sep 16 & IMACS & $u'g'r'i'z'$ & $720; 240; 240; 270; 480$  & 0\farcs67 \\
              & 2007 Sep 22 & PANIC & $JHK_s$      & $810; 1080; 1080$          & 1\farcs06 \\
\rxnine       & 2007 Feb 13 & MagIC & $g'r'i'z'$   & $240; 240; 300; 480$       & 0\farcs70 \\
              & 2007 Feb 14 & MagIC & $u'$         & $720$                      & 0\farcs63 \\
              & 2007 Apr 7  & PANIC & $JHK_s$      & $540; 405; 405$            & 0\farcs72 \\
\sdssnine     & 2007 Feb 13 & MagIC & $u'g'r'i'z'$ & $720; 240; 240; 180; 480$  & 0\farcs61 \\
              & 2007 Apr 7  & PANIC & $JHK_s$      & $540; 540; 540$            & 0\farcs64 \\
\heoneone     & 2007 Feb 14 & MagIC & $g'r'i'z'$   & $240; 180; 120; 240$       & 0\farcs49 \\
              & 2007 Apr 7  & PANIC & $JHK_s$      & $450; 450; 450$            & 0\farcs45 \\
\pgoneone     & 2008 Feb 1  & MagIC & $g'r'i'z'$   & $120; 120; 150; 240$       & 0\farcs55 \\
              & 2008 Feb 2  & MagIC & $u'$         & $720$                      & 0\farcs87 \\
              & 2008 Feb 4  & PANIC & $JHK_s$      & $540; 540; 540$            & 0\farcs56 \\
\rxoneone     & 2007 Feb 13 & MagIC & $u'g'r'i'z'$ & $480; 120; 120; 120; 240$  & 0\farcs65 \\
              & 2007 Apr 6  & PANIC & $JHK_s$      & $486; 405; 405$            & 0\farcs76 \\
\sdssoneone   & 2007 Feb 13 & MagIC & $u'g'r'i'z'$ & $1080; 480; 480; 360; 960$ & 0\farcs75 \\
              & 2007 Apr 8  & PANIC & $JHK_s$      & $540; 540; 540$            & 0\farcs47 \\
\sdssonethree & 2008 Feb 1  & MagIC & $u'g'r'i'z'$ & $360; 120; 120; 150; 240$  & 0\farcs53 \\
              & 2008 Feb 3  & PANIC & $JHK_s$      & $540; 540; 540$            & 0\farcs68 \\
\wfitwosix    & 2008 May 13 & MagIC & $u'g'r'i'z'$ & $720; 240; 240; 360; 480$  & 0\farcs54 \\
              & 2008 May 12 & PANIC & $JHK_s$      & $135; 135; 90$             & 0\farcs43 \\
\wfithree     & 2007 Jun 15 & MagIC & $u'g'r'i'z'$ & $720; 240; 120; 300; 480$  & 0\farcs88 \\
              & 2007 Jul 4  & PANIC & $JHK_s$      & $135; 135; 90$             & 0\farcs51
\enddata
\end{deluxetable*}

\begin{figure}
  \centering
  \scalebox{-1}[1]{\includegraphics[width=0.8in]{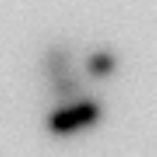}}
  \scalebox{-1}[1]{\includegraphics[width=0.8in]{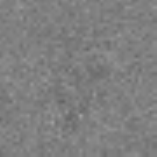}}\hspace{.025in}
  \scalebox{-1}[1]{\includegraphics[width=0.8in]{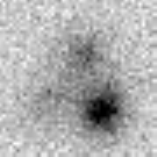}}
  \scalebox{-1}[1]{\includegraphics[width=0.8in]{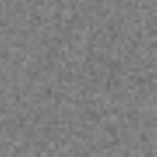}}\\[.05in]
  \scalebox{-1}[1]{\includegraphics[width=0.8in]{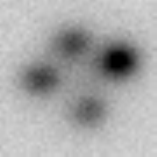}}
  \scalebox{-1}[1]{\includegraphics[width=0.8in]{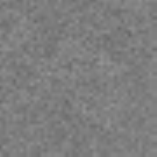}}\hspace{.025in}
  \scalebox{-1}[1]{\includegraphics[width=0.8in,origin=c,angle=270]{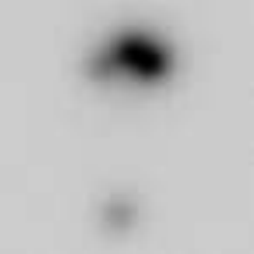}}
  \scalebox{-1}[1]{\includegraphics[width=0.8in,origin=c,angle=270]{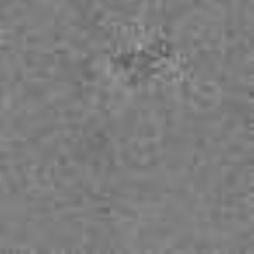}}\\[.05in]
  \scalebox{-1}[1]{\includegraphics[width=0.8in,origin=c,angle=270]{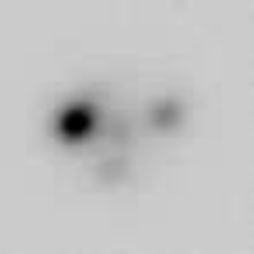}}
  \scalebox{-1}[1]{\includegraphics[width=0.8in,origin=c,angle=270]{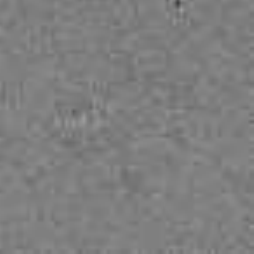}}\hspace{.025in}
  \scalebox{-1}[1]{\includegraphics[width=0.8in,origin=c,angle=270]{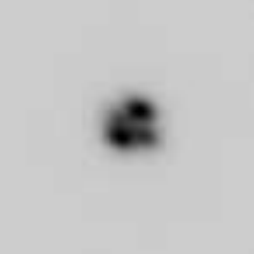}}
  \scalebox{-1}[1]{\includegraphics[width=0.8in,origin=c,angle=270]{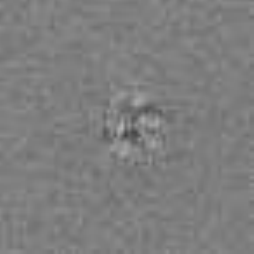}}\\[.05in]
  \scalebox{-1}[1]{\includegraphics[width=0.8in,origin=c,angle=180]{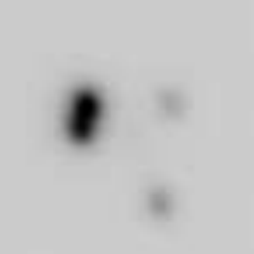}}
  \scalebox{-1}[1]{\includegraphics[width=0.8in,origin=c,angle=180]{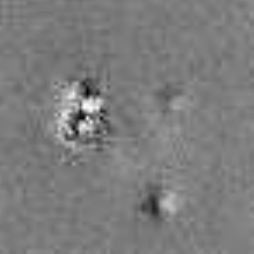}}\hspace{.025in}
  \scalebox{-1}[1]{\includegraphics[width=0.8in,origin=c,angle=270]{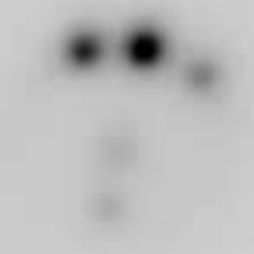}}
  \scalebox{-1}[1]{\includegraphics[width=0.8in,origin=c,angle=270]{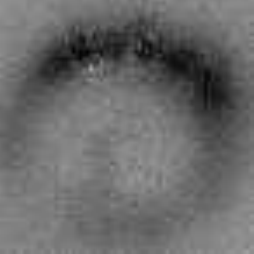}}\\[.05in]
  \scalebox{-1}[1]{\includegraphics[width=0.8in,origin=c,angle=270]{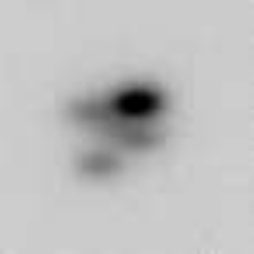}}
  \scalebox{-1}[1]{\includegraphics[width=0.8in,origin=c,angle=270]{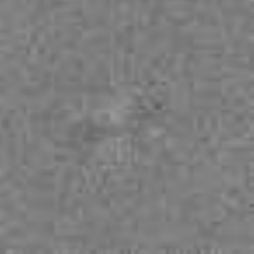}}\hspace{.025in}
  \scalebox{-1}[1]{\includegraphics[width=0.8in,origin=c,angle=180]{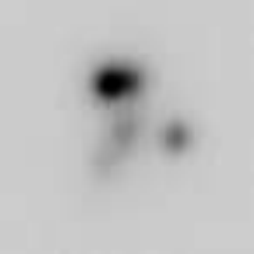}}
  \scalebox{-1}[1]{\includegraphics[width=0.8in,origin=c,angle=180]{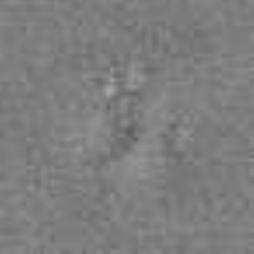}}\\[.05in]
  \scalebox{-1}[1]{\includegraphics[width=0.8in]{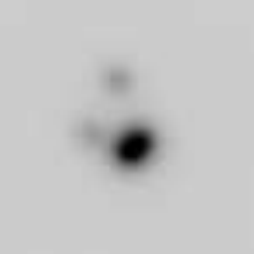}}
  \scalebox{-1}[1]{\includegraphics[width=0.8in]{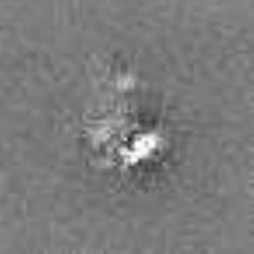}}\hspace{.025in}
  \scalebox{-1}[1]{\includegraphics[width=0.8in,origin=c,angle=90]{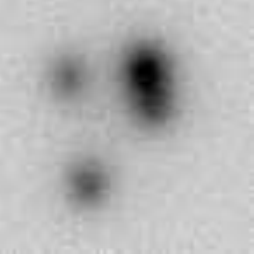}}
  \scalebox{-1}[1]{\includegraphics[width=0.8in,origin=c,angle=90]{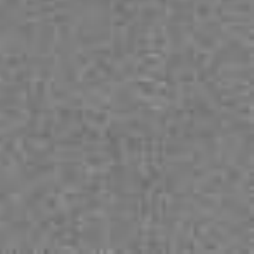}}
  \caption{Postage-stamp images and residuals of lenses in the $i'$
    band. From top left, they are \hetwo, \mgfour, \hefour, \rxnine,
    \sdssnine, \heoneone, \pgoneone, \rxoneone, \sdssoneone,
    \sdssonethree, \wfitwosix, and \wfithree. Each frame is $5''$ on a
    side, with north up and east left. The gray-scale stretch on the
    residual images is set to $\pm20\sigma$ of the sky level. For
    \rxoneone, we show the residuals of the ``chi-by-eye'' fit, where
    the flux ratios are set by hand to account for the bright Einstein
    ring.}
  \label{fig:stamps}
\end{figure}

Between 2007 February and 2008 May, we undertook an optical observing
campaign to obtain multi-band, contemporaneous images of our sample of
lenses. At IR wavelengths, we used the $J$, $H$, and $K_s$
filters with Persson's Auxiliary Nasmyth Infrared Camera (PANIC), on
the 6.5 m Baade telescope at Las Campanas Observatory. In the
optical, we used the Sloan $u'g'r'i'z'$ filters with the Raymon and
Beverly Sackler Magellan Instant Camera (MagIC), which was on the
neighboring Clay telescope when we began our campaign, but was moved
to Baade while the observations were still underway. This change of
location allowed us to use both instruments during the same observing
run, and even during the same night. MagIC and PANIC have fields
of view 2\farcm4 and 2\farcm1 on a side, respectively, large enough to
include stars appropriate for use as PSF templates. The instruments'
pixel scales are 0\farcs069 and 0\farcs125 pixels, which more than
adequately sample the PSF. The details of the observations are listed
in Table~\ref{tab:optobs}.

During our 2007 September observing run, MagIC was offline for an
upgrade, so we instead used the Inamori Magellan Areal Camera System
(IMACS) in its imaging mode. With its $f/4$ camera, the instrument has a
pixel scale of 0\farcs11 and a 15\farcm5 field of view, but in order
to reduce the readout time we used only a subraster 2\farcm2 on a
side. We used this instrument for our observations of \hetwo, \mgfour,
and \hefour.

The images were corrected for bias and dark current, flattened, and
cleaned of cosmic rays using standard techniques. Where multiple
exposures were obtained, they were stacked, resulting in a single
image per filter per lensed quasar. Postage-stamp $i'$-band images of
the 12 lenses in our sample are provided in Figure~\ref{fig:stamps},
together with the residuals from the PSF fitting described in
Section~\ref{sec:photometry}.

Because of the small separations between the quasar images in our
sample, we only observed under the best atmospheric seeing conditions,
typically below 0\farcs7. The excellent image quality at the Magellan
telescopes was crucial to the success of our survey. The $i'$- and
$J$-band seeing is reported in Table~\ref{tab:optobs}, which outlines
the optical/IR observations obtained for this work. The seeing in the
other filters was roughly consistent with the rule of thumb:
$\mathrm{FWHM} \propto \lambda^{-1/5}$.

Because we are primarily interested in the flux {\em ratios} of lensed
quasars (indeed, we cannot measure absolute magnifications), we did
not, in general, obtain images of standard stars for the purpose of
calibrating our photometry. In the case of \sdssonethree, however, we
did observe standard stars in the optical bands, so in
Section~\ref{sec:photometry} we report calibrated photometry for this
lens. We also estimate the photometric zero point in the $J$ and $K_s$
bands, based on images taken the same night which contain several
calibrated sources from the Two-Micron All Sky Survey \citep[2MASS;
][]{Skrutskie:2006p1163}.

Despite the high quality of our data, careful PSF subtraction was
necessary to disentangle the compact clumps of point sources and lens
galaxies that make up our sample. For each image, we used a nonlinear
least-squares fitting algorithm to simultaneously fit the positions
and relative fluxes of the four quasar components and the lens galaxy.

For lenses with well-separated components, we performed a simultaneous
fit to the images in all eight bands, with the relative positions of the
quasar components and the lens galaxy allowed to vary, but constrained
to be consistent across all eight filters. The flux ratios of the quasar
components were free to vary in each filter independently. The
best-fit positions were consistent in every case with the \textit{HST}
positions from the CASTLES survey. In cases where merging pairs or
triplets of images would have caused strong correlations between
positions and flux ratios, we fixed the relative positions of the
quasar components and the lensing galaxy to the CASTLES positions. We
modeled the quasar images using nearby relatively bright stars as
empirical PSF templates.

In the $r'$ and redder bands, we did not use least-squares
minimization to determine the flux ratios of \rxoneone. Because of
the bright Einstein ring in this system, this strategy would have
over-subtracted the quasar components. Instead we fixed all of the
quasar fluxes to values that resulted in residuals that looked like an
unbroken Einstein ring. Although it was completely ad hoc, we think that
this ``chi by eye'' technique gave flux ratios less affected by
systematic errors. Section~\ref{sec:optuncertainty} describes our
method for estimating the uncertainty of these flux ratios.

In all cases we modeled the lensing galaxy as a two-dimensional
pseudo-Gaussian function \citep{Schechter:1993p1342}, with a
full-width at half-max as a fixed parameter. The widths were chosen
using a trial-and-error technique, examining the residuals by
eye. (Because of the frequent presence of a faint Einstein ring due to
the quasar's host galaxy, least-squares minimization often
overestimated the width.) Because the lens galaxy was nearly always
faint compared to the quasar images, the goodness of the fit was
insensitive to this approximation. The galaxies were mostly round, but
for the few exceptions we fixed the axis ratio and position angle by
hand, as we did the width.

The fitted fluxes of the lens galaxies are uncalibrated, and therefore
of little interest; we do not report them. \sdssonethree\ is the
exception; because it was discovered relatively recently, and because
observing conditions were photometric on the night it was observed, we
observed standard stars in the optical bands, for the purpose of
calibrating its fluxes. The calibrated magnitudes of this lens galaxy
are reported in Table~\ref{tab:1330phot}. The optical calibration
relies on aperture photometry of the standard stars \mbox{SA 107-351}
(for $u'$) and \mbox{G163-50} (for $g'r'i'z'$), and places the fluxes
on the Sloan $u'g'r'i'z'$ photometric system, which closely
approximates the monochromatic AB system \citep{Smith:2002p2121}. In
the $H$ and $K_s$ bands, we made use of a field containing several
2MASS sources for photometric calibration; we performed aperture
photometry on them and compared their measured magnitudes to those
reported in the NASA/IPAC Infrared Science
Archive\footnote{\url{http://irsa.ipac.caltech.edu}} to obtain
photometric zero points. This places our calibrated $J$ and $K_s$
fluxes on the 2MASS photometric system
\citep{Cohen:2003p1090}. In the optical bands, used the atmospheric
extinction coefficients of \citet{Smith:2007p0} to correct for
differences in airmass between \sdssonethree\ and the standard stars;
these corrections were all less than 0.04 mag.

The lack of \textit{HST} positions for \sdssonethree, combined with the
relative brightness and elliptical shape of the lensing galaxy, made
its fit a special challenge. In order to get the shape of the galaxy
right, we adopted an iterative approach. First, we subtracted a rough
model of the lens galaxy, fixing its parameters to reasonable
guesses. We then fit the positions and fluxes of the quasar
components, subtracting them from the image. We added back our rough
galaxy model to the residual image and fit a single pseudo-Gaussian to
the resulting galaxy image, with all parameters allowed to vary. The
best-fit FWHM of the galaxy was $1\farcs0$ along the major axis; the
axis ratio and position angle were $0.67$ and $24\fdg5$ east of
north. Finally, we fit the original image again, fixing the galaxy
shape parameters to these best-fit values, in order to find the quasar
positions and fluxes. The results are consistent with those of
\citet{Oguri:2008p1973}, but have higher precision; they are reported
in Tables~\ref{tab:1330astro} (astrometry) and \ref{tab:1330phot}
(photometry).

For each lens, the transformation between pixel coordinates and sky
coordinates was determined by registering our $i'$- and $J$-band images
to the \mbox{USNO-B} astrometric catalog. On average, there were $\sim
10$--$15$ objects for each fit. The pixel scale and rotation angle thus
determined for $i'$ and $J$ were assumed to apply to the remaining
MagIC/IMACS data and PANIC data for that lens,
respectively. We tested this assumption for a few cases by fitting
images in other bands to the catalog and found it to be justified.

Relative photometry for all the lenses in our sample except
\sdssonethree\ is listed in Table~\ref{tab:optdata}. We also list the
calculated rms deviation of the magnitudes in each filter from those
predicted by the lens models. This number is an indication of how
anomalous the flux ratios are. For comparison with the optical flux
ratios, we also list the X-ray ratios, in magnitudes.


\section{Uncertainty Estimation}
\label{sec:uncertainty}

Aside from statistical measurement uncertainties in the flux ratios,
systematic errors arise from several sources. At optical and
IR wavelengths, quasar emission lines, quasar variability, or
contamination from a lens galaxy or an Einstein ring may
contribute. In X-rays, quasar variability is a factor, and delays
between X-ray and optical/IR observations combine with
microlensing variability to contribute additional systematic
uncertainty.

\begin{deluxetable}{ccc}
\tablewidth{0pt} 
\tablecaption{Relative Astrometry for \sdssonethreelong
  \label{tab:1330astro}}
\tablehead{\colhead{Component} & \colhead{$x$} & \colhead{$y$}}
\startdata
A & $-1.26 \pm 0.01$ & $-1.17 \pm 0.01$       \\
B & $-0.84 \pm 0.01$ & $-1.19 \pm 0.01$       \\
C & $\equiv 0$        & $\equiv 0$              \\
D & $-1.49 \pm 0.01$ & $\phm{-}0.44 \pm 0.01$ \\
G & $-1.04 \pm 0.02$ & $-0.20 \pm 0.02$    
\enddata
\tablecomments{All positions are in arcseconds. Positive $x$ and $y$
point west and north, respectively.}
\end{deluxetable}

\subsection{Optical Uncertainties}
\label{sec:optuncertainty}

\begin{deluxetable*}{cccccc}
\tabletypesize{\small}
\tablewidth{0pt}
\tablecaption{Photometry for \sdssonethreelong
  \label{tab:1330phot}}
\tablehead{
  \colhead{Filter}& 
  \colhead{A $-$ C} & 
  \colhead{B $-$ C} & 
  \colhead{C} & 
  \colhead{D $-$ C} &
  \colhead{G}
}
\startdata
$u'$  & $-0.75\pm0.01\pm0.05$ & $+0.27\pm0.02\pm0.05$ & $19.69\pm0.06\pm0.01$ & $+1.70\pm0.02\pm0.05$ & \nodata \\
$g'$  & $-0.79\pm0.01\pm0.05$ & $+0.17\pm0.01\pm0.05$ & $20.99\pm0.06\pm0.01$ & $+1.78\pm0.02\pm0.05$ & $25.08\pm0.11$ \\
$r'$  & $-0.86\pm0.01\pm0.05$ & $-0.18\pm0.01\pm0.05$ & $20.57\pm0.06\pm0.01$ & $+1.63\pm0.02\pm0.05$ & $23.48\pm0.11$ \\
$i'$  & $-0.90\pm0.01\pm0.00$ & $-0.13\pm0.01\pm0.00$ & $21.07\pm0.06\pm0.01$ & $+1.71\pm0.02\pm0.01$ & $23.26\pm0.10$ \\
$z'$  & $-0.88\pm0.01\pm0.00$ & $-0.12\pm0.02\pm0.00$ & $21.55\pm0.06\pm0.01$ & $+1.95\pm0.05\pm0.01$ & $23.46\pm0.10$ \\
$J$   & $-0.85\pm0.02\pm0.05$ & $-0.18\pm0.03\pm0.05$ & $18.88\pm0.06\pm0.01$ & $+1.33\pm0.06\pm0.05$ & $18.94\pm0.10$ \\
$H$   & $-0.84\pm0.03\pm0.05$ & $-0.24\pm0.03\pm0.05$ & \nodata               & $+1.54\pm0.09\pm0.05$ & \nodata \\
$K_s$ & $-0.89\pm0.04\pm0.00$ & $-0.36\pm0.04\pm0.00$ & $17.78\pm0.06\pm0.01$ & $+2.29\pm0.30\pm0.01$ & $17.13\pm0.10$
\enddata
\tablecomments{Optical magnitudes for image C and lens galaxy G are
  calibrated to the Sloan $u'g'r'i'z'$ system, while $J$ and $K_s$
  magnitudes are calibrated to the 2MASS system. Formal statistical
  uncertainties are reported first, followed by estimated systematic
  uncertainties. No systematics are reported for the lens galaxy G.}
\end{deluxetable*}

Quasar emission lines are thought to come from a region too large to
be strongly affected by microlensing\footnote{Some degree of broad
  line emission microlensing has been observed, however
  \citep[e.g.,][]{Keeton:2006p1}.}
\citep{Schneider:1990p42}. Therefore the presence of emission line
flux in our broadband measurements will cause errors if we assume we
are measuring only the continuum from the accretion disk. The strength
and effect of these errors is very difficult to predict. We allocated
a 0.05 mag systematic uncertainty in our flux ratios for filters into
which one of the following lines has been redshifted: C\textsc{iv},
C\textsc{iii}$]$, Mg\textsc{ii}, H$\beta$, or H$\alpha$. (This is very
  roughly the percentage of the broadband flux taken up by one of
  these lines, in general.)  Occasionally, Ly$\alpha$ falls in a
  filter, or there are two emission lines present; in these cases, we
  allocated 0.1 mag of uncertainty.

The multiple images of a strongly lensed quasar arrive with relative
delays of hours to weeks because of the different paths taken by their
light. Quasar variability can conspire with these time delays to mimic
flux ratio anomalies. We do not expect this to be a very strong
effect, because quad lenses tend to have short time delays, especially
between close pairs and triplets where the most interesting anomalies
tend to happen. In order to quantify the effect, we extrapolated the
empirical quasar variability structure function in Figure~18 of
\citet{deVries:2005p615} using a power law: \mbox{$\log S(\tau) = 0.8
  + 0.65\log \tau$}. This gave us an estimate for the standard
deviation $S$ of the quasar brightness in magnitudes as a function of
time delay $\tau$ in years.

Time delays have been measured for five of the lenses in our
sample. For \hefour, \rxnine, \pgoneone, \rxoneone, and \wfithree, we
used, respectively, the time delays reported by
\citet{Kochanek:2006p47}, \citet{Hjorth:2002pL11},
\citet{Barkana:1997p21}, \citet{Morgan:2006p5321}, and
\citet{Vuissoz:2008p481}. For the remaining lenses we used the time
delays predicted by our lens models (see
Section~\ref{sec:models}). The lens redshifts of \heoneone\ and
\wfitwosix\ are unknown, so we estimated them both to be
$z_L\approx0.7$ using the method of \citet{Kochanek:1992p1}. The
resulting quasar variability uncertainties were only significant for a
few quasar images (e.g., image D in \rxoneone\ or \rxnine).

Finally, the formal measurement uncertainties for a few flux ratios
were clear underestimates, or (in one case) simply did not
exist. \hetwo\ provides an example, as its component D is blended in
our images with a bright companion to the main lens galaxy only
0\farcs{}4 away. For this flux ratio, we allocated an extra error
equal to the change in measured brightness if 25\% of the galaxy light
were attributed to the quasar image. This ranged from 0.1 mag in $g'$
to 1.6 mag in $K_s$. The flux ratios of \rxoneone\ have no formal
uncertainties because of the ad hoc method we used to estimate them
(see Section~\ref{sec:photometry}). We conservatively estimated their
uncertainties as half the difference between our estimates and those
of the poorly performing chi-square fit. In addition, we allocated 0.5
mag of extra uncertainty to images C and D of \sdssnine\ in the
$u'$ band, which were only marginally detected, and 0.05 mag to
images A1 and A2 of \wfitwosix\ in the $H$ band, whose flux ratios
depended slightly on our choice of PSF comparison star.

Table~\ref{tab:optdata} shows the formal statistical uncertainty of
each flux ratio, followed by the systematic uncertainty, defined as the
quadratic sum of that due to emission lines, intrinsic variability,
and other systematics, as described above.

\subsection{X-ray Uncertainties}
\label{sec:xrayuncertainty}

Like the optical flux ratios, the X-ray ratios had measurement
uncertainties due to measurement noise and the blending of close
pairs; both errors were generally larger because the X-ray
observations have fewer photons and a broader PSF. There were
also contributions from intrinsic quasar variability, as in the
optical case. The X-ray ratios had no systematic errors due to
emission lines, but they did have errors due to microlensing
variability because they were not generally measured contemporaneously
with the other wavelengths.

Crucial to our analysis is the assumption that the arrangement of the
source and the microlenses is the same for all wavelengths. But when
observations are not contemporaneous, the source and the microlenses
have the opportunity to reconfigure themselves. To estimate the
magnitude of this effect, we again used a structure function; this
time it was not empirical, but was derived from microlensing
magnification patterns (see our description of these patterns in
Section~\ref{sec:magmaps}). For each image, we chose the appropriate
pattern (with the stellar mass fraction $k_*$ set to 0.1; see
Section~\ref{sec:magmaps}) and ran 1000 tracks across it, in random
directions, measuring the variance in the logarithm of the
magnification as a function of distance moved.

The conversion to a structure function (with a time delay on the
abscissa instead of a distance) required an estimate of the transverse
speed of the source relative to the lens. We added four velocity
components in quadrature: the velocity dispersion of the stars in the
lens galaxy (as estimated by our lens model), the tangential component
of the velocity of the Sun relative to the rest frame of the cosmic
microwave background (CMB), and the peculiar velocities of the quasar
and the lens galaxy \citep[as estimated using Equation~(14.10)
  of][]{Peebles:1980p0}:
\begin{align}
v_{\perp}^2 = \notag
&2\left(\frac{\sigma_L}{(1 + z_L)}
\frac{D_\mathrm{OS}}{D_\mathrm{OL}}\right)^2
+\left(v_\mathrm{CMB} \sin{\alpha}
\frac{D_\mathrm{LS}}{D_\mathrm{OL}}\right)^2 \\
+&2\left(\frac{\sigma_\mathrm{pec}}{(1 + z_S)^{3/2}} 
\frac{f(z_S)}{f(0)}\right)^2
+2\left(\frac{\sigma_\mathrm{pec}}{(1 + z_L)^{3/2}} 
\frac{f(z_L)}{f(0)}\frac{D_\mathrm{OS}}{D_\mathrm{OL}}\right)^2 ~.
\end{align}
Here $z_S$ and $z_L$ are the redshifts of the source and the lens,
respectively, and $D_\mathrm{OL}$, $D_\mathrm{OS}$, and
$D_\mathrm{LS}$ are angular diameter distances from observer to lens,
observer to source, and lens to source. We used these distances to
project all velocities to the source plane. The angle $\alpha$ is
measured between the Sun's velocity with respect to the CMB rest frame
and the line of sight to the lens, and we set $v_\mathrm{CMB}$ to 370
km s$^{-1}$ \citep{Lineweaver:1996p38}. We followed
\citet{Kochanek:2004p58} in setting the present-day velocity
dispersion of galaxies $\sigma_\mathrm{pec}$ to 235 km s$^{-1}$. The
function $f(z)$ is the cosmological growth factor; we approximate it
as $f \propto \Omega_M(z)^{0.6}$. The stellar velocity dispersion
$\sigma_L$ of the lens we estimate from the monopole component $b$ of
our lens model (see Section~\ref{sec:models}) using the relation
\citep[e.g.,][]{Kochanek:2006p91}
\begin{equation}
b = 4\pi \frac{\sigma_L^2}{c^2} \frac{D_\mathrm{LS}}{D_\mathrm{OS}} ~.
\end{equation}
All velocities are corrected for cosmological time dilation. The
factors of two convert one-dimensional velocity dispersions to two
dimensions. Multiplying the estimated source speed $v_{\perp}$ by the
delay between our optical/IR observations and their X-ray
counterparts, we determined the distance traveled by the source along
the caustic pattern, and from the structure function read off the
predicted error in our X-ray flux ratio.

For each X-ray flux ratio, we added in quadrature the uncertainty
contributions from measurement errors (including extra uncertainty for
image A2 of \wfitwosix\ because of its degeneracy with the PSF
width), intrinsic quasar variability, and microlensing
variability. These uncertainties are given in
Table~\ref{tab:optdata}. In general, the largest contributions were
systematic uncertainties from microlensing variability, though the
statistical measurement errors were also substantial in some cases. As
in the optical case, the uncertainty due to quasar variability was
relatively insignificant.

\addtocounter{table}{1}
\begin{deluxetable*}{lcccccccccccc}
\tablewidth{0pt}
\tablecaption{Lens Model Parameters
  \label{tab:modelparams}}
\tablehead{
  \colhead{} &
  \multicolumn{3}{c}{Primary Lens} &
  \colhead{} &
  \multicolumn{3}{c}{Secondary Lens} &
  \colhead{} &
  \multicolumn{4}{c}{Magnification\tnm{a}}\\
  \cline{2-4}
  \cline{6-8}
  \cline{10-13}
  \colhead{Quasar} & 
  \colhead{$b$} & 
  \colhead{$\gamma$} & 
  \colhead{$\phi_{\gamma}$\tnm{b}} & 
  \colhead{} &
  \colhead{$b_2$} & 
  \colhead{$x_2$\tnm{c}} & 
  \colhead{$y_2$\tnm{c}} & 
  \colhead{} &
  \colhead{HM} & 
  \colhead{HS} & 
  \colhead{LM} & 
  \colhead{LS}
}
\startdata
\hetwo        & $0\farcs87$ & $0.11\phn$                            & $-60\fdg0$           && $0\farcs33$ & $\phn\mbox{$-$}0.283$ & $+0.974$ && $+9.42$ & $-9.65$ & $+4.95$ & $-1.35$ \\
\mgfour       & $1\farcs14$ & $0.11\phn$                            & $+74\fdg6$           && $0\farcs12$ & $\phn\mbox{$-$}0.385$ & $+1.457$ && $+22.9$ & $-24.2$ & $+6.23$ & $-3.11$ \\
\hefour       & $1\farcs20$ & $0.078$                               & $-13\fdg8$           && \nodata     & \nodata               & \nodata  && $+7.49$ & $-7.90$ & $+7.14$ & $-4.73$ \\
\rxnine       & $0\farcs97$ & $0.27\phn$                            & $\phn\mbox{+}7\fdg2$ && $0\farcs24$ & $\phn\mbox{$-$}0.754$ & $+0.665$ && $+11.0$ & $-5.96$ & $+1.97$ & $-4.99$ \\
\sdssnine     & $0\farcs87$ & $0.063$                               & $+84\fdg8$           && \nodata     & \nodata               & \nodata  && $+14.9$ & $-13.0$ & $+6.62$ & $-6.55$ \\
\heoneone     & $0\farcs33$ & $0.040$                               & $+37\fdg7$           && \nodata     & \nodata               & \nodata  && $+15.8$ & $-16.7$ & $+12.6$ & $-9.59$ \\
\pgoneone     & $1\farcs03$ & \nodata                               & \nodata              && $2\farcs57$ & $-10.866$             & $-5.300$ && $+19.7$ & $-18.9$ & $+5.09$ & $-3.37$ \\
\rxoneone     & $1\farcs86$ & $0.16\phn$                            & $-73\fdg6$           && \nodata     & \nodata               & \nodata  && $+13.2$ & $-22.7$ & $+12.6$ & $-1.05$ \\
\sdssoneone   & $0\farcs67$ & $0.10\phn$                            & $+32\fdg6$           && \nodata     & \nodata               & \nodata  && $+7.17$ & $-6.68$ & $+5.17$ & $-3.64$ \\
\sdssonethree & $0\farcs94$ & $0.16$\makebox[0pt][l]{\tnm{d}}$\phn$ & $-32\fdg2$\tnm{d}    && \nodata     & \nodata               & \nodata  && $+27.1$ & $-27.2$ & $+8.41$ & $-5.50$ \\
\wfitwosix    & $0\farcs66$ & $0.11\phn$                            & $-90\fdg0$           && \nodata     & \nodata               & \nodata  && $+13.7$ & $-11.5$ & $+3.78$ & $-4.01$ \\
\wfithree     & $1\farcs07$ & $0.11\phn$                            & $+36\fdg0$           && $0\farcs25$ & $\phn\mbox{$+$}0.229$ & $+2.020$ && $+6.04$ & $-3.80$ & $+3.88$ & $-2.46$  
\enddata
\tablenotetext{a}{HS: highly magnified saddle point; HM: highly
  magnified minimum; LS: less magnified saddle point; LM: less
  magnified minimum.}
\tablenotetext{b}{Position angle of external shear $\gamma$. All
angles are measured in degrees east of north.}
\tablenotetext{c}{Fixed position of secondary galaxy, relative to main
  lensing galaxy, in arcseconds. Allowed to vary radially in the case
  of \pgoneone; see the text. The positive directions of $x$ and $y$
  are west and north, respectively.}
\tablenotetext{d}{Ellipticity and position angle of the lens galaxy for the SIE model of \sdssonethree.}
\end{deluxetable*}

\section{Lens Models}
\label{sec:models}

We used the {\it lensmodel} program of \citet{Keeton:2001p2340} to
create parametric models of each lens. The models were constrained by
the positions of the four lensed images, and that of the lensing
galaxy, a total of 10 constraints. We did not use fluxes for
constraints, since most of our lenses suffer from flux ratio
anomalies. Nor did we use time delays. The positions for the quasar
images and lens galaxies came from the CASTLES survey.

Our default model consisted of a singular isothermal sphere for the
lensing galaxy, with a quadrupole component of the potential provided
by a constant external shear. With the position of the lens fixed,
this model had five free parameters: the monopole strength of the
lens, the magnitude and direction of the shear, and the position of
the source.

In some cases, such as \hefour, this simple model fit the image
positions well. But in several cases the $\chi^2$ goodness of fit was
poor enough to warrant further complexity in the model. In these
cases, we made changes to the model motivated by the appearance of the
lens galaxy.

The lens galaxy of \hetwo\ has a prominent companion, located close to
image D. We modeled this companion as a second isothermal sphere,
fixing its position to its \textit{HST} measured value but allowing its
mass to vary. Despite only adding one free parameter to the model,
this addition improved the fit considerably. We followed a similar
strategy for \mgfour, \rxnine, and \wfithree, each of which displays a
faint smudge in its \textit{HST} image which is arguably a satellite to
the lens galaxy. Adding secondary lenses at the positions of the
smudges improved the goodness of fit to acceptable levels.

The lens galaxy in \pgoneone\ does not have a nearby companion, but is
a member of a small galaxy group centered to the southwest of the
lens. Following the example of \citet{Schechter:1997pL85}, we
explicitly modeled the group as a second isothermal sphere for this
lens. We parameterized its position using polar coordinates, and
allowed its mass and distance from the main galaxy to vary while
fixing its position angle to that of the brightest galaxy in the
group. We did not include an external shear in this fit.

Finally, the lens galaxy in \sdssonethree\ displays significant
ellipticity. In this case we used an isothermal ellipsoid instead of
including an external shear. We allowed the ellipticity and position
angle to vary, along with the galaxy position, for a total of seven
free parameters. Since there are no \textit{HST} data for this lens, we
used our measured image positions for constraints (see
Section~\ref{sec:photometry} and
Table~\ref{tab:1330astro}). \sdssonethree\ was the only lens system
where an isothermal ellipsoid made for a better fit than an isothermal
sphere with external shear.

The salient features of our best-fit models are listed in
Table~\ref{tab:modelparams}. We used the components of the
magnification tensor predicted by these models (namely, the local
convergence $\kappa$ and shear $\gamma$) to generate the microlensing
magnification patterns described in Section~\ref{sec:magmaps}. We also
used the predicted scalar magnifications to estimate the
microlensing-free flux ratios, and assumed that any discrepancies
between the predicted and observed flux ratios was due to
microlensing. We believe that this is a good assumption for our sample
of lenses because there is not strong evidence for other potential
explanations. For example, differential extinction is unlikely to be a
factor in these early-type lens galaxies \citep{Kochanek:2004p69};
this is supported by the general lack of strong hydrogen absorption in
X-rays \citep{Pooley:2007p19}. Likewise, the chromatic variations we
see in the flux ratios argues against millilensing by dark matter
substructure. In the case of \mgfour, however, mid-infrared flux
ratios measured by \citet{Minezaki:2009p610} disagree with the
predicted ratios. Since the dusty torus region where this light
originates is too large to be affected by microlensing, it is likely
that millilensing is affecting this lens. In this case, we adopted the
reported mid-IR flux ratios as the ``baseline'' rather than the model
ratios, attributing any optical departures from them to microlensing.

The predicted flux ratios are model-dependent to some extent; this
systematic error is difficult to quantify, but ought to be small
compared to the other uncertainties. \citet{Dalal:2002p25} suggest
error bars of 10\%; \citet{Keeton:2003p138} call this ``quite
conservative.'' We estimated the model uncertainties in the flux
ratios as 0.05 mag for well-separated quasar images, or 0.03
mag for members of a close pair or triplet, since these
configurations have model-independent asymptotic flux ratios
\citep[][and references therein]{Keeton:2003p138, Keeton:2005p35}. We
added this uncertainty in quadrature to those listed in
Table~\ref{tab:optdata}.


\section{Bayesian Source Size Estimation}
\label{sec:analysis}

The effect of a finite source size is to smooth out the temporal
variations caused by microlensing. The movement of a large source
across a network of microlensing magnification caustics is equivalent
to a point source moving across a version of the caustic map that has
been smoothed by convolution with the surface brightness profile of
the source. Since microlensing variability is stronger in X-rays than
at optical wavelengths, it is thought that the X-rays come from a very
compact region, whereas the optical source is much larger
\citep{Pooley:2007p19, Pooley:2009p1892, Morgan:2008p755,
  Dai:2010p278}. For single-epoch observations like ours, microlensing
magnifications are more naturally treated as flux ratio anomalies,
where the flux ratios of the quasar images fail to match those
predicted by lens models, than as variability. We expect the flux
ratio anomalies to be stronger in X-rays than at optical wavelengths,
and stronger in blue light than in red light, most of the
time. However, different regions of the magnification maps have
different responses to a change in the source size. Indeed, in some
regions (e.g., at the center of an astroid caustic), convolution by a
large source profile will {\em increase} the
anomaly. \citet{Poindexter:2008p34} saw an example of this in the
doubly lensed quasar HE\,$1104-1805$.

Single-epoch measurements are much less expensive than long-term
monitoring campaigns, especially with multiple wavelengths. But they
have the disadvantage that they are unable to constrain the local
characteristics of the microlensing magnification map, leading to a
degeneracy between the natural range of possible magnifications and
source size effects. A strong anomaly is evidence for a relatively
small source, but a weak or absent anomaly could be due to a large
source or to the chance location of a small source on a relatively
calm portion of the magnification map. \pgoneone\ is a real-life
example of this: for several years its optical A2/A1 ratio maintained
a low-level anomaly that was revealed to be due to a large source size
when the same ratio was found to be extremely low in X-rays
\citep{Pooley:2006p67}. Our analysis method follows
\citet{Pooley:2007p19} in using X-ray flux ratios to break this
degeneracy, but in a more quantitative way.

A similar uncertainty exists when using size estimates at a range of
wavelengths to estimate the wavelength dependence of the size of
quasar accretion disks. Because of the unknown response of the
microlensing magnification to changes in source size, a single-epoch
observation of a single lensed quasar cannot tell us much about this
dependence. We stress, however, that it does allow for estimates with
meaningful error bars. The way to decrease this uncertainty is to
sample a greater fraction of the caustic map, either via time-series
observations of a single lens or by observing a sample of lenses. We
have adopted the latter approach.

We have developed a quantitative Bayesian method for estimating the
angular size of quasar accretion disks at a range of optical and
IR wavelengths using our measured optical/IR and X-ray flux
ratios. This approach relates the posterior probability distribution
for the source size to the likelihood of observing the reported X-ray
flux ratios, assumed to originate in a very compact source,
simultaneously with the observed optical/IR ratios from a larger
source (see Equation~(\ref{eqn:bayes1})). The bulk of this section is
devoted to the calculation of these likelihood distributions.

\subsection{Microlensing Magnification Patterns}
\label{sec:magmaps}

We simulated the microlensing of a finite-size source using
magnification patterns created using the inverse ray-tracing software
of \citet{Wambsganss:1990p407}. Each pattern corresponds to a specific
quasar image, and was constructed using the local convergence
$\kappa_\mathrm{tot}$ and shear $\gamma$ specified by our lens models
(see Section~\ref{sec:models}). The convergence (which is proportional
to the surface mass density) we divided into $\kappa_c$ due to
continuously distributed matter (e.g., dark matter) and $\kappa_*$ due
to stellar-mass point lenses. For each quasar image we created 12
patterns, varying the surface mass fraction in stars $k_* \equiv
\kappa_*/\kappa_\mathrm{tot}$ from 1.47\% to 100\%, in logarithmic
steps.

The point mass microlenses were scattered randomly across the
microlensing patterns, with masses drawn at random from a fixed mass
function. The primary effect of the mass function chosen for the
simulated microlenses is to set the physical size of the microlens
Einstein radius, the natural scale for microlensing
simulations. Different mass functions also lead to differently shaped
magnification histograms, but this effect is less important. We were
not able to make an estimate of the mean mass $\langle m \rangle$
using our single-epoch measurements, so we adopted a fixed broken
power-law mass function nearly identical to the well-known
\citet{Kroupa:2001p231} initial mass function. Between $0.08\,M_\odot$
and $0.5\,M_\odot$, its logarithmic slope was $-1.8$; above
$0.5\,M_\odot$ it steepened to $-2.7$. We cut off the mass function at
$1.5\,M_\odot$, because the stellar populations in these early-type
lens galaxies are typically old. With this mass function, the average
microlens mass $\langle m \rangle$ was $0.247\,M_\odot$. With this
fixed mean mass came a fixed transformation between the angular sizes
we measure (in units of Einstein radii) and physical sizes at the
distance of the quasar (in units of cm); any adjustment to it will
scale our measured disk sizes according to $\langle m \rangle^{1/2}$.

Each pattern encodes the deviation in the magnification of its quasar
image from that produced by a smooth mass distribution, as a function
of the position of the source. They are 2000 pixels on a side, and
represent a square region of the source plane with a side length 20
times the projection of the Einstein radius of a solar-mass star in
the lens galaxy. This distance is roughly $5\times10^{17}$ cm for the
lenses in our sample, though the exact numbers depend weakly on the
lens and source redshifts. With these dimensions, the patterns are
large enough that their histograms are insensitive to the exact
configuration of the stars. The pixel size is ~$2.5\times10^{14}$ cm,
or a few gravitational radii for a $10^9\,M_\odot$ black hole. This is
much smaller than the size of the optical accretion disk and is
comparable to the source size in X-rays \citep{Morgan:2008p755,
Dai:2010p278}.

We modeled sources of finite size by convolving the magnification
patterns with circular Gaussian kernels of varying half-light
radii. Though a Gaussian is not a very physical choice for a source
profile, it has been shown \citep{Mortonson:2005p594,Congdon:2007p263}
that the half-light radius is much more important when simulating
finite sources than are the details of the radial profile. This
convolution blurs the patterns and reduces their dynamic range,
causing their histograms to become narrower. The result of our efforts
was a family of patterns for each quasar image, parameterized by
stellar mass fraction and source size.

The local surface mass fraction in stars as a function of position
within a galaxy is poorly constrained. \citet{Pooley:2009p1892} used
X-ray flux ratios of \pgoneone\ to estimate the stellar mass fraction
at the locations where the quasar images intersect the lens
galaxy. Because the quasar appears to be compact in X-rays
\citep{Pooley:2007p19,Chartas:2009p174}, they were able to safely
neglect the finite source size, and found a probability distribution
for the stellar mass fraction $k_*$ that was peaked near 10\%. Since a
detailed study of stellar mass fractions is beyond the scope of this
work, we marginalized over this parameter using the results found by
\citet{Pooley:2009p1892}(see Sections~\ref{sec:1dhist} and
\ref{sec:2dhist}).

\subsection{Size Estimation Using Optical Ratios}
\label{sec:1dhist}

We used a simple Bayesian method to calculate the posterior
probability distribution for the half-light radius of the source
$r_{1/2}$ at each wavelength. Assuming a uniform prior for the source
size, the probability is
\begin{equation}
\label{eqn:bayes1}
p\left(r_{1/2}|\{m_i\}\right) \propto \mathcal{L}\left(\{m_i\}|r_{1/2}\right)~,
\end{equation}
where $\{m_i\}$ is the set of quasar image magnitudes, $i$ runs from
$1$ to $N_\mathrm{im} = 4$, and $\mathcal{L}$ is the likelihood of
having observed the magnitudes $\{m_i\}$ given a source half-light
radius. Each magnitude $m_i$ may be expressed as the sum of the
unmagnified quasar magnitude $m_S$, macro- and micro-magnification terms
$\mu_i$ and $\Delta \mu_i$, and measurement noise $n_i$:
\begin{equation}
m_i = m_S + \mu_i + \Delta \mu_i + n_i~.
\end{equation}
The source magnitude $m_S$ is not observable, and because we did not
flux-calibrate our measurements, it is in uncalibrated units. Since we
ultimately marginalized over $m_S$, these limitations are unimportant.

We computed the right-hand side of Equation~(\ref{eqn:bayes1}) by
comparing the measured image fluxes $m_i$ to histograms of the
magnification patterns described in Section~\ref{sec:magmaps}. Suitably
normalized, each histogram was an estimate of the distribution of
microlensing magnifications:
\begin{equation}
\label{eqn:hist1d}
p_i\left(\Delta \mu_i | r_{1/2}, k_*\right) ~,
\end{equation}
where the pattern sourcing the histogram was created using the stellar
mass fraction $k_*$ and was convolved with a source with radius
$r_{1/2}$.

\begin{figure}
  \centering
  \includegraphics[width=0.45\textwidth]{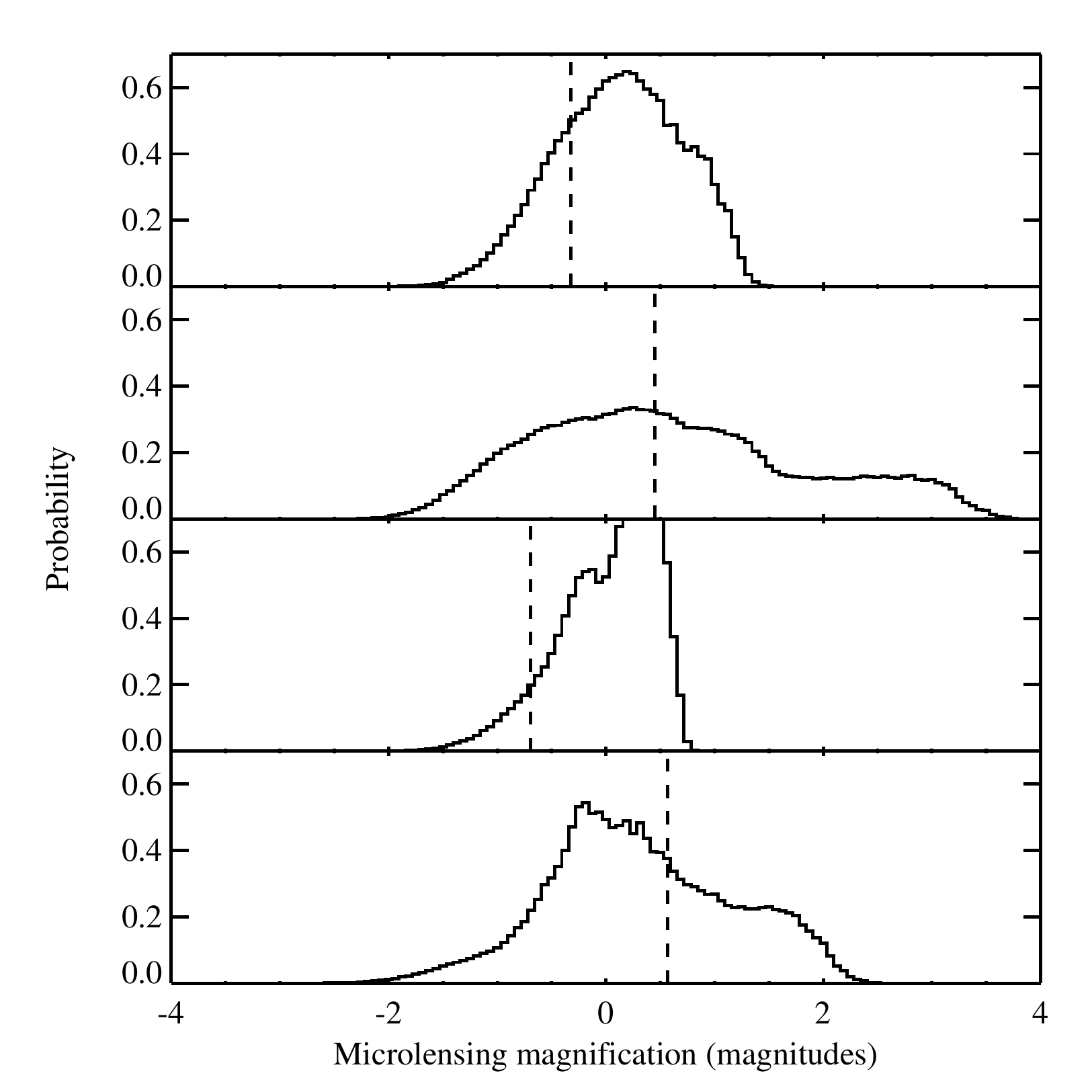}
  \caption{Histograms derived from magnification patterns for the four
    images A, B, C, and D of \sdssonethree\ (top to bottom), with a
    source radius 3.2\% of a solar-mass Einstein radius and a stellar
    fraction of 32\%. Flux ratios from our $i'$ band data have been
    overplotted in dotted lines. The abscissa denotes deviation from
    the predicted flux ratios; if the dotted lines all fell at zero,
    the lens would have no flux ratio anomaly. Magnifications are in
    magnitudes, so positive numbers denote demagnification. The
    histograms of the negative parity images (B and D) are notably
    wider than those of the positive parity images. The same is true
    of the more highly magnified images (A and B).}
  \label{fig:4hist}
\end{figure}

Figure~\ref{fig:4hist} shows an example of four such magnification
histograms, one for each image in \sdssonethree, for a particular
source size and stellar mass fraction. The abscissa of this plot
denotes the microlensing deviation $\Delta \mu$ in magnitudes. Atop
the histograms have been plotted vertical lines, indicating the
observed deviation from the model in the $i'$ band. In the case of no
microlensing, the lines would be co-located at zero. The presence of
{\em four} vertical lines misleadingly implies more information than
is available; in fact, we have not measured the four image
magnifications, but only their three ratios. Thus, the average
position of the four lines is unknown. In Figure~\ref{fig:4hist}, we
have set it to zero, but in reality the four lines are free to slide
from side to side in formation. Each horizontal position corresponds
to a distinct value of the quasar's unknown magnitude $m_S$.

We shifted each histogram horizontally to obtain $p_i(\Delta \mu_i -
m_i + \mu_i) = p_i(-m_S - n_i)$.  These shifts brought the four vertical
lines in Figure~\ref{fig:4hist} together. Because the shifts depended
on the measured fluxes, we marginalized over the measurement errors:
\begin{equation}
\label{eqn:convolve1}
p_i(-m_S) = \frac{1}{\sqrt{2\pi}\sigma_i} \int dn_i 
\exp{\left(\frac{-n_i^2}{2\sigma_i^2}\right)} p_i(-m_S - n_i) ~.
\end{equation}
This amounted to a convolution of the shifted histograms with the
normal distributions implied by the measurement uncertainties
$\sigma_i$. These shifted and convolved histograms were probability
density functions of the quasar magnitude $m_S$. We multiplied them
together and integrated their product to find the likelihood of
measuring all the observed flux ratios, independent of the quasar
flux:
\begin{equation}
\mathcal{L}\left(\{m_i\}|r_{1/2},k_*\right) = \int dm_S \phi(m_S)
\left[\prod_{i=1}^{N_\mathrm{im}} p_i(-m_S)\right] ~,
\end{equation}
where $N_\mathrm{im} = 4$ and $\phi(m_S)$ is the prior on the luminosity
of the quasar. For this luminosity function we used a power law of
slope $-2.95$, as estimated for bright quasars by
\citet{Richards:2006p2766}. Our use of a scale-free luminosity
function obviated any difficulties due to the uncalibrated quasar
flux.

Finally, we marginalized over the stellar mass fraction $k_*$ to find
the right-hand side of Equation~(\ref{eqn:bayes1}):
\begin{equation}
\label{eqn:kmarg1}
\mathcal{L}\left(\{m_i\}|r_{1/2}\right) = \int dk_* p(k_*)
\mathcal{L}\left(\{m_i\}|r_{1/2},k_* \right)~.
\end{equation}
For our prior $p(k_*)$ on the stellar mass fraction we interpolated
the estimated probability distribution for its value given in Figure 6
of \citet{Pooley:2009p1892}. It should be noted that although we have
written the marginalizations in
Equations~(\ref{eqn:convolve1})--(\ref{eqn:kmarg1}) as integrals for
clarity, they were actually calculated as discrete sums.

We calculated this relative likelihood for 31 half-light radii, spaced
logarithmically between $0.01r_\mathrm{Ein}$ and $10r_\mathrm{Ein}$,
where $r_\mathrm{Ein}$ is the Einstein radius of a solar-mass
microlens. We let the posterior probability of the source size equal
this likelihood, implicitly adopting a logarithmic prior on the source
size (between these limits; zero elsewhere). For completeness, we also
used a linear prior; this simply involved multiplying the likelihood
distribution by a power law of unit slope.

\begin{figure}
  \centering
  \includegraphics[width=0.45\textwidth]{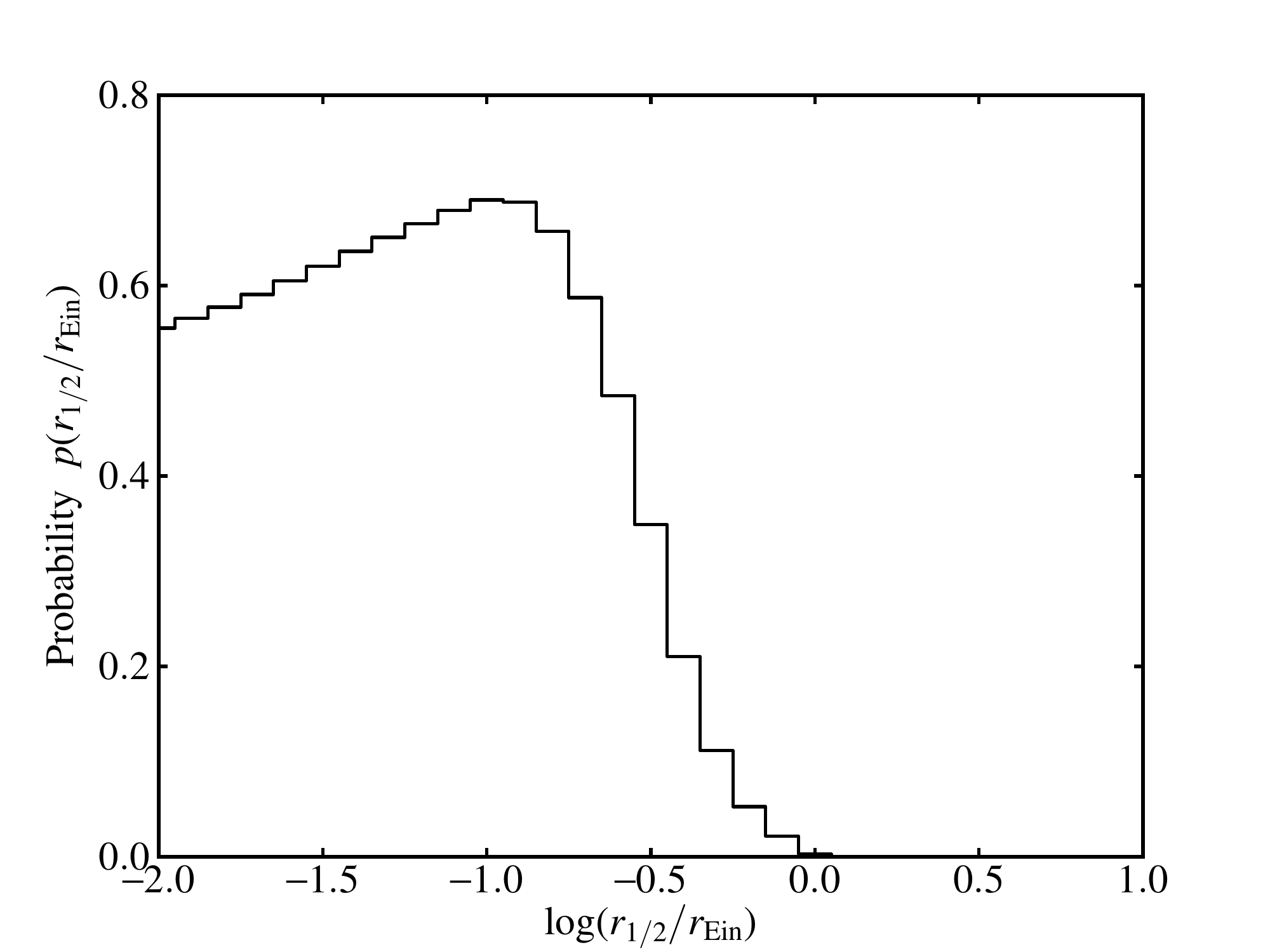}
  \caption{Posterior probability distribution for the size of
    \sdssonethree, using the $i'$ band data. The histograms were
    convolved with the measurement uncertainties, and the quasar
    luminosity function was used as a weighting factor when
    integrating the product of the histograms. A logarithmic prior was
    used for the source size.}
  \label{fig:examplelikeli1}
\end{figure}

Figure~\ref{fig:examplelikeli1} shows the posterior distribution
resulting from the use of this technique on the $i'$-band flux ratios
of \sdssonethree. This figure demonstrates the ability of this method
to rule out large source sizes (e.g., greater than an Einstein
radius). The decrease in probability to zero occurs when the growing
source size causes two of the shifted histograms to become narrow
enough that they no longer overlap at all; at this point the
likelihood is zero. Measurement errors widen the histograms via the
convolution in Equation~(\ref{eqn:convolve1}), causing the probability
distribution to widen as well. There is no corresponding decrease in
probability for small source radii. This is to be expected, since the
optical flux ratio anomalies simply indicate that the source is
undergoing microlensing. To set a lower limit on the size, we must
discern why the observed anomalies are not stronger than they are---is
the caustic map being smoothed out by a large source, or is the source
simply lying in a non-perturbed region?  Though some size
discrimination is possible by comparing the posterior distributions
derived from blue filters versus red filters, a more effective method
to rule out very small sizes is to compare the optical flux ratios to
those in X-rays.

\subsection{Size Estimation Using Optical and X-ray Ratios}
\label{sec:2dhist}

\sdssonethree, the lens we chose for the examples in the previous
section, is unique among our sample in that it has never been observed
in X-rays. But for the remaining lenses in our sample there is at
least one measurement of the X-ray flux ratios. Since the microlensing
variability is in general stronger in X-rays than at optical
wavelengths, the X-rays must come from a quite compact region
\citep{Pooley:2007p19, Pooley:2009p1892, Dai:2010p278}. Any decrease
in the anomalies in the optical and IR must be due to the finite
size of the source at these wavelengths. Comparing the optical flux
ratios to X-ray ratios gives us the ability to set a lower limit on
the size of the optical-emitting accretion disk.

\begin{figure}
  \centering
  \includegraphics[width=0.45\textwidth]{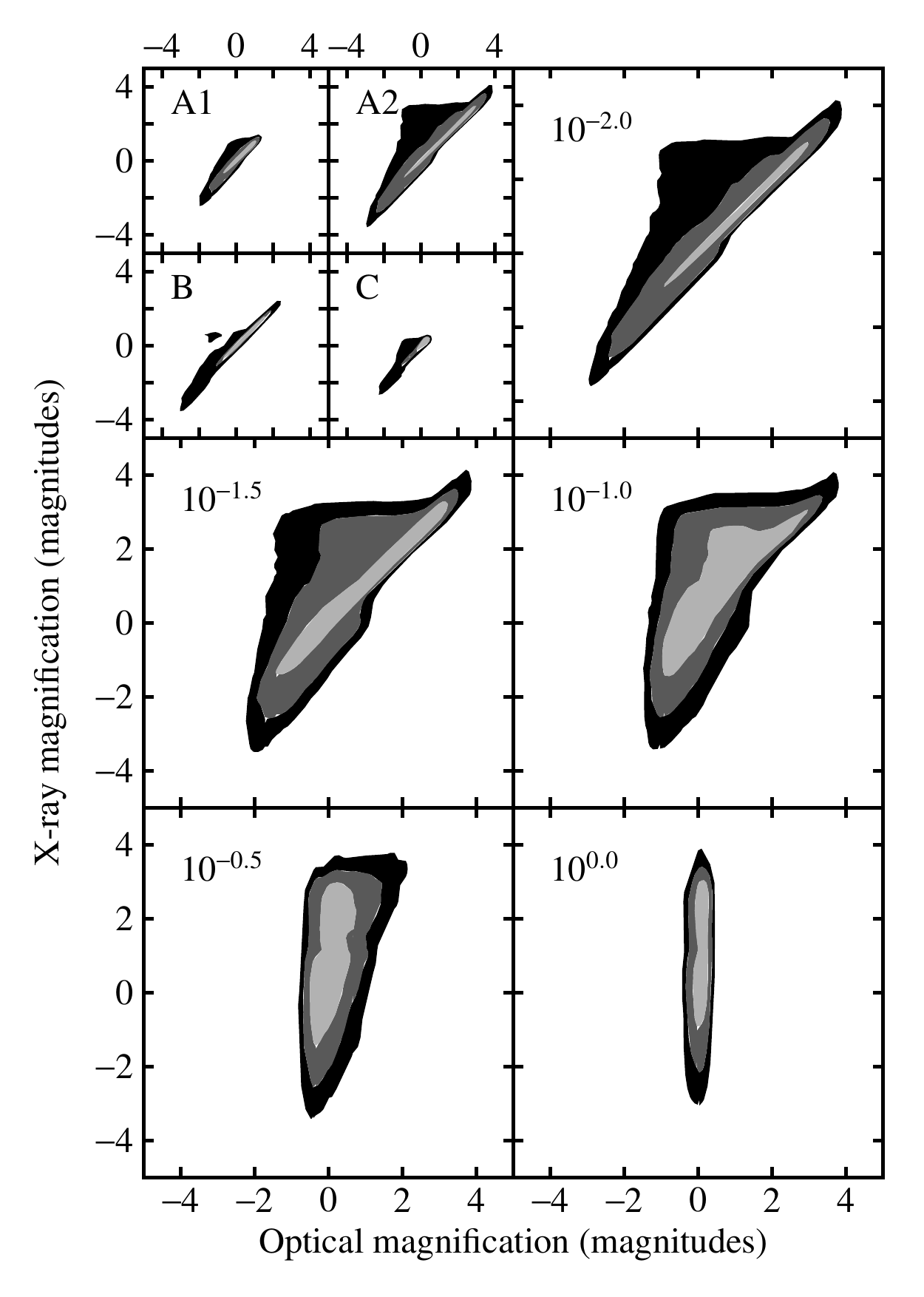}
  \caption{$1\sigma$, $2\sigma$, and $3\sigma$ contours of the
    two-dimensional microlensing magnification histograms for the four
    images of \pgoneone, after the magnification map has been
    convolved with Gaussian source profiles with varying optical half
    light radii. In the upper left panel, histograms for all four
    images are shown, with an optical source size of $10^{-2}
    r_\mathrm{Ein}$. The other panels show the evolution of the
    histogram of image A2 as the optical size increases; the half
    light radii are labeled in the upper left of each panel, in units
    of $r_\mathrm{Ein}$. As the source gets larger, the correlation
    between X-ray and optical magnifications decreases, and the
    histogram narrows in the optical direction. For even larger source
    sizes, the histograms get very narrow.}
  \label{fig:hist2d}
\end{figure}

We again wished to find the value of the right-hand side of
Equation~(\ref{eqn:bayes1}), but this time, we wanted the likelihood of
observing \emph{both} the X-ray and optical flux ratios
simultaneously, given a source size. Because the X-rays are thought to
originate from a very compact region, we assumed that the X-ray ratios were
drawn from the original (unconvolved) magnification pattern. Then, for
each image, we constructed a family of two-dimensional histograms:
\begin{equation}
p_i\left(\Delta \mu_i^X, \Delta \mu_i^O | r_{1/2}, k_*\right) ~,
\end{equation}
where $\Delta \mu_i^O$ and $\Delta \mu_i^X$ indicate values taken from
the convolved (optical) and unconvolved (X-ray) versions of the same
pattern. These histograms, after appropriate scaling, are joint
probability distributions for the optical and X-ray microlensing
magnifications; they are the two-dimensional generalization of the
histogram in Equation~(\ref{eqn:hist1d}). Figure~\ref{fig:hist2d}
shows two-dimensional histograms corresponding to the images of
\pgoneone, with the X-ray magnifications drawn from the original
pattern and the optical magnifications drawn from the same pattern
smoothed by a source of varying half-light radii. Histograms for all
four images are shown for a single source size ($0.01
r_\mathrm{Ein}$), and the histogram for the highly magnified saddle
point image A2 is shown for a variety of source sizes. The maps used
for these histograms all had a stellar mass fraction $k_* = 0.1$.

We shifted the histograms along the optical and X-ray axes in a manner
analogous to the one-dimensional case, using the measured flux ratios
in X-rays and in the chosen optical/IR filter. For very small source
sizes, the two patterns from which each histogram was constructed were
nearly identical, so the histograms showed a significant correlation
between $\Delta \mu^O$ and $\Delta \mu^X$. Thus, for a very small
source size to be assigned any significant probability, the optical
and X-ray shifts had to be the same. As the optical source size
increased, the histograms' extent in the $\Delta \mu^O$ direction
decreased, as did the correlation between the magnification in the two
bands, until for very large source sizes the histograms were nearly
aligned with the $\Delta \mu^X$ axis. In this case, the optical shifts
had to be nearly zero for the probability to be appreciable (see
Figure~\ref{fig:hist2d}).

In the one-dimensional case, we propagated the uncertainties in the
measured flux ratios by convolving our shifted histograms with
Gaussians representing the uncertainties. In this two-dimensional
context, convolution (e.g., with a two-dimensional Gaussian) gave us
insufficient information. Since the source size likelihood in every
filter depended on the same X-ray flux ratios, the X-ray uncertainties
led to correlated errors in the size estimates at different
wavelengths. So we instead performed a Monte Carlo integration in
order to propagate the full covariance matrix through the analysis
described in the next paragraph. We describe the error propagation in
the paragraph that follows.

We multiplied the shifted histograms and marginalized over
the unknown X-ray and optical source fluxes to find the likelihood of
simultaneously measuring the observed X-ray and optical fluxes:
\begin{align}
\mathcal{L}\left(\{m_i^X,m_i^O\}|r_{1/2},k_*\right) = &\int dm_S^X dm_S^O 
\phi\left(m_S^X,m_S^O\right) \nonumber \\
&\times \left[\prod_{i=1}^{N_\mathrm{im}} p_i\left(-m_S^X,-m_S^O\right)\right] ~.
\end{align}
In this case, the weighting function $\phi(m_S^X,m_S^O)$ incorporated
a prior both on the luminosity function of quasars and on the
correlation between their X-ray and optical fluxes. The latter is
often parameterized using $\alpha_{OX} \equiv 0.3838 \log
(L_\mathrm{2\,keV}/L_{2500\text{\AA}})$. \citet{Gibson:2008p773} find
that $\alpha_{OX} = -0.217\log(L_{2500\text{\AA}})$ plus a constant
offset, with a scatter of 0.1. This implies that $m_S^X = 0.4346m_S^O$
plus a constant offset, with a scatter of 0.65 mag. So we constructed
a band centered on a line with a slope of 0.4346 in the
$(m_S^X,m_S^O)$ plane, and with a Gaussian cross-section. In order to
be conservative, we doubled the observed scatter and set the width of
the Gaussian to correspond to $\sigma = 1.3$ magnitudes. We adjusted
the constant offset so that the line passed through the mean of the
product histogram. We multiplied this function by a power law of slope
$-2.95$ (as in the one-dimensional case) in the direction along the
line of correlation to represent the luminosity function. We note that
the slope of the luminosity function of X-ray selected bright quasars
is very similar at $-2.8$ \citep{Silverman:2008p118}. As before, we
repeated this technique using magnification patterns representing a
range of stellar mass fractions, and marginalized over this parameter
using a weighted average of the resulting likelihood distributions
(see Equation~(\ref{eqn:kmarg1})).

The Monte Carlo integration which allowed us to propagate the
measurement uncertainties into the likelihood distributions consisted
of repeating the process of shifting, multiplying, and integrating
under the histograms for 1000 sets of flux ratios drawn from Gaussian
distributions implied by the measurement uncertainties. This gave us
1000 distributions for the disk size in every filter. From these we
took the most likely values (the modes) and constructed the covariance
matrix $C_{ij} = \mathrm{Cov}(r_{1/2,i},r_{1/2,j})$, where $i$ and $j$
refer to optical/IR filters. The diagonal elements of the matrix
are simply the variances of the half-light radii due to measurement
uncertainties. To this matrix we added a second, diagonal matrix
$\mathcal{S}_{ij}$, which expressed the ``intrinsic'' width of the
first of our 1000 likelihood distributions, the one derived from the
reported flux ratios. This intrinsic width arose not from measurement
uncertainties, but simply from the widths of the magnification
histograms. The diagonal elements of $\mathcal{S}$ were the squares of
the halved 68\% confidence intervals of these distributions, and the
remaining elements were zero. We describe in the next section how we
used the inverse of the sum of these covariance matrices
\begin{equation}
\label{eqn:weightmatrix}
\mathcal{W} = (C + \mathcal{S})^{-1}
\end{equation}
as a weight matrix in a chi-square fit.

\begin{figure}
  \centering
  \includegraphics[width=0.45\textwidth]{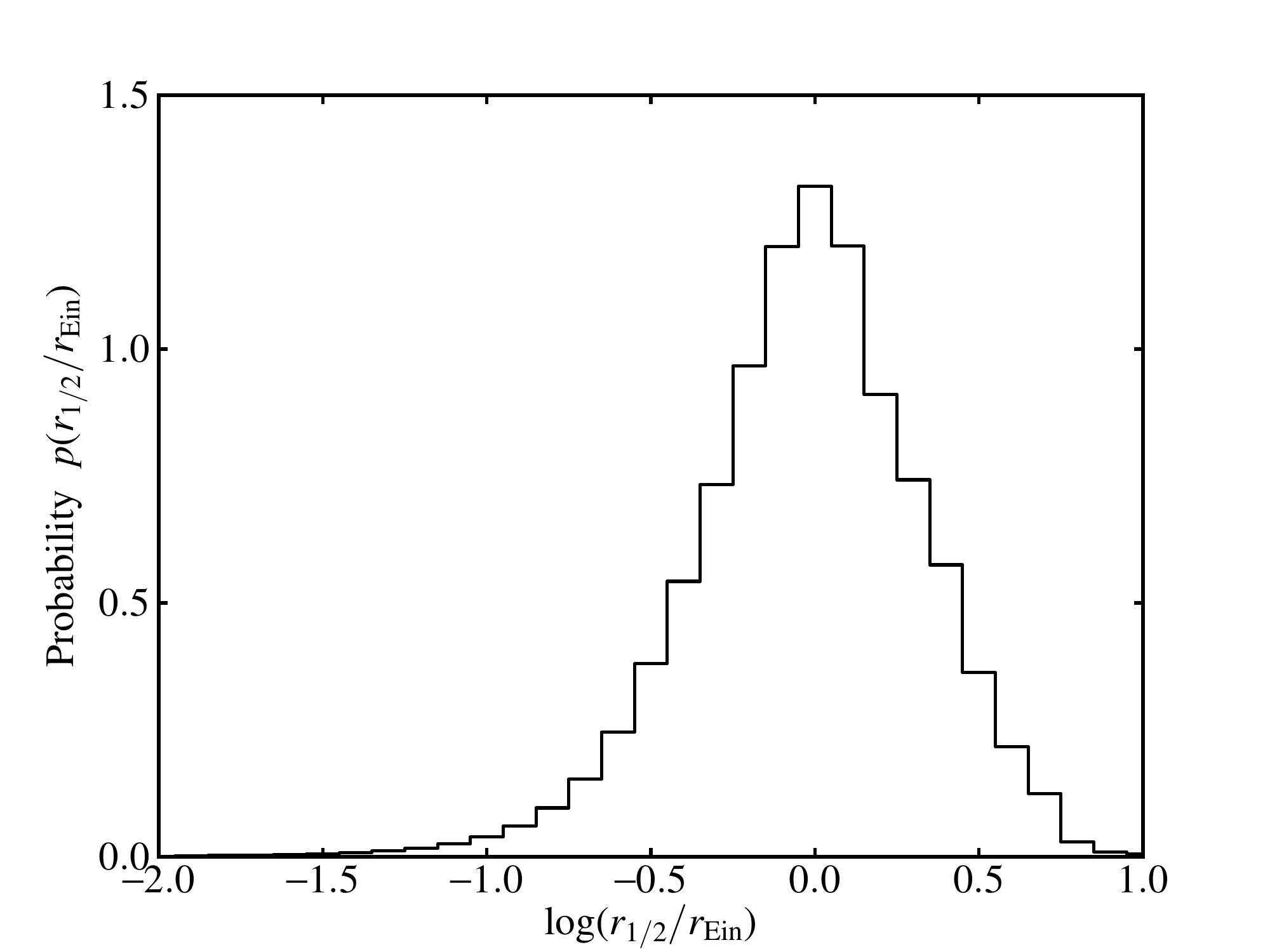}
  \caption{Posterior probability distribution for the size of
    \pgoneone\ in the $i'$ band, resulting from considering both
    $i'$-band and X-ray flux ratios. The Monte Carlo method described
    in Section \protect{\ref{sec:2dhist}} was used to account for
    measurement uncertainties. A logarithmic prior was used for the
    source size.}
  \label{fig:examplelikeli2}
\end{figure}

The fact that the optical flux ratios demonstrate microlensing
magnifications allows us to place upper limits on the size of the
optical disk, since a very large disk will be unaffected by
microlensing. Likewise, the fact that the X-ray flux ratios are
different from (typically, larger than) the optical ratios allows us
to place lower limits. For small optical sizes, the differing flux
ratios cause the highly correlated histograms to be shifted in a
direction different from that of the correlation, lowering the
likelihood (see Figure~\ref{fig:hist2d}). The posterior probability
distribution for the $i'$-band radius of the accretion disk of
\pgoneone, determined in this manner, is shown in
Figure~\ref{fig:examplelikeli2}, and shows a central peak, with a
decrease in probability to the left due to the interaction of the
X-ray and optical flux ratios, and a decrease to the right due
primarily to the optical flux ratios alone.

We tested the accuracy of our analysis method by applying it to
artificial flux ratios generated using the predictions of the thin
disk model. We calculated the half-light radii that this model
predicts for \pgoneone\ in each filter, then convolved the
magnification patterns appropriate for the four images of this lens
(with stellar mass fraction $k_* = 0.1$) with source profiles of these
radii. Then we drew microlensing magnifications from 30 random
locations in these patterns. We drew X-ray ratios from the same
locations in the corresponding non-smoothed patterns. We ran our
analysis on these 30 sets of simulated flux ratios, with uncertainties
equal to those assigned to the real \pgoneone\ data, to check if we
recovered the input temperature profile. We fit a line to the
recovered sizes to determine the logarithmic slope and normalization
of the wavelength-dependent half-light radius, using the procedure
described in Section~\ref{sec:results}. On average, the best-fit
overall normalization $\log(r_{1/2}/\mathrm{cm})$ was 16.1, with a
scatter of 0.2 dex, compared to the input value of 15.7. The average
output slope was 1.2, with a scatter of 0.1, versus the input value of
1.33. The general agreement of the recovered parameter values with the
input values indicates that the analysis method produces sensible
results. The small offset between the input and output normalizations
indicates that our analysis may overestimate the sizes slightly.

\begin{figure}
  \centering
  \includegraphics[width=0.45\textwidth]{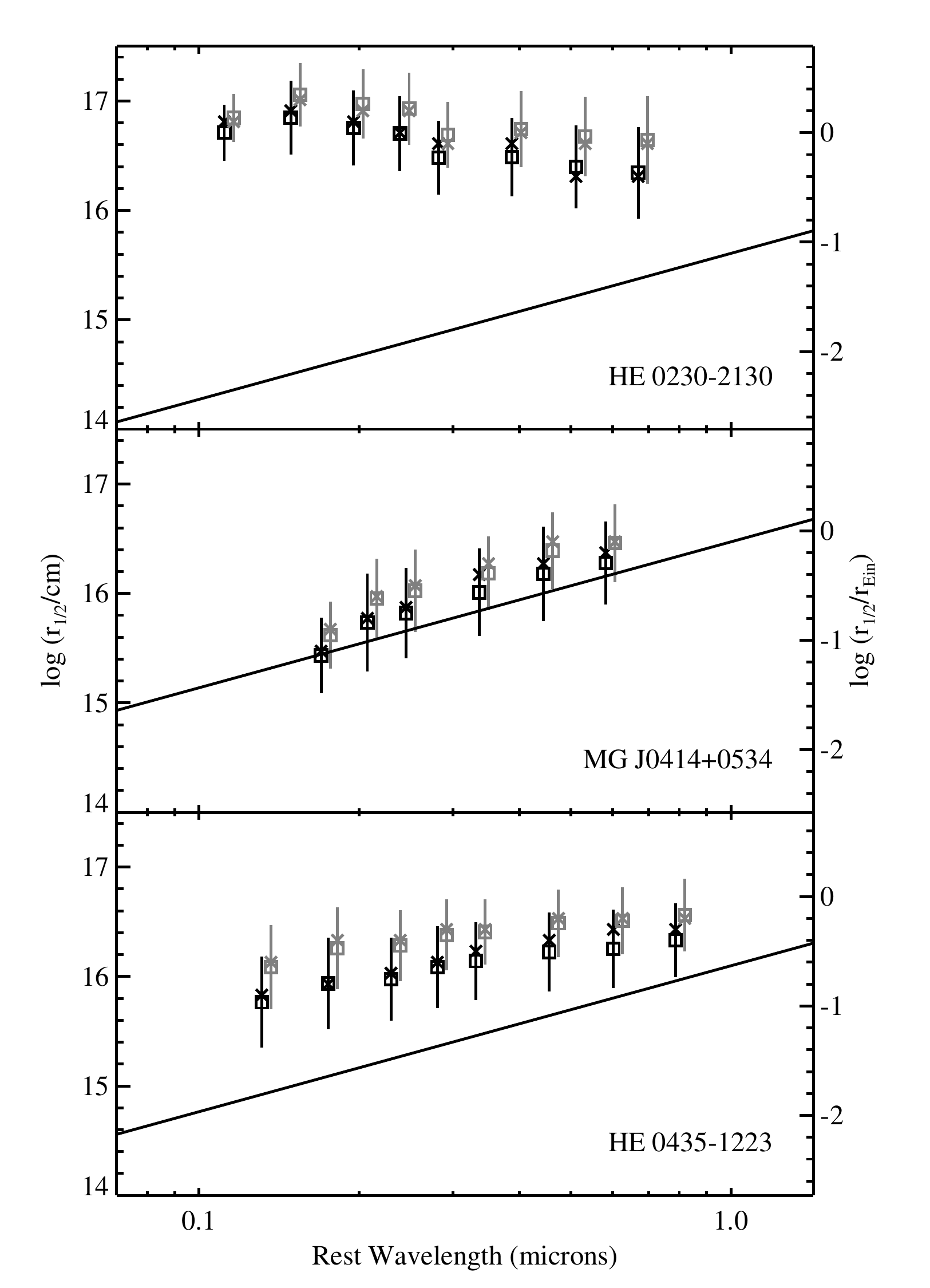}
  \caption{Medians (squares) and modes (crosses) of the probability
  distributions for the half-light source radius. The error bars
  correspond to the quadratic sum of the intrinsic width of the
  distributions and the scatter from the Monte Carlo propagation of
  measurement uncertainties. Black (gray) points indicate a
  logarithmic (linear) prior. The solid line is the prediction for the
  half-light radius (in cm) of the standard thin disk model, for the
  estimated black hole mass.}
  \label{fig:size1}
\end{figure}


\section{Results}
\label{sec:results}

\begin{figure}
  \centering
  \includegraphics[width=0.45\textwidth]{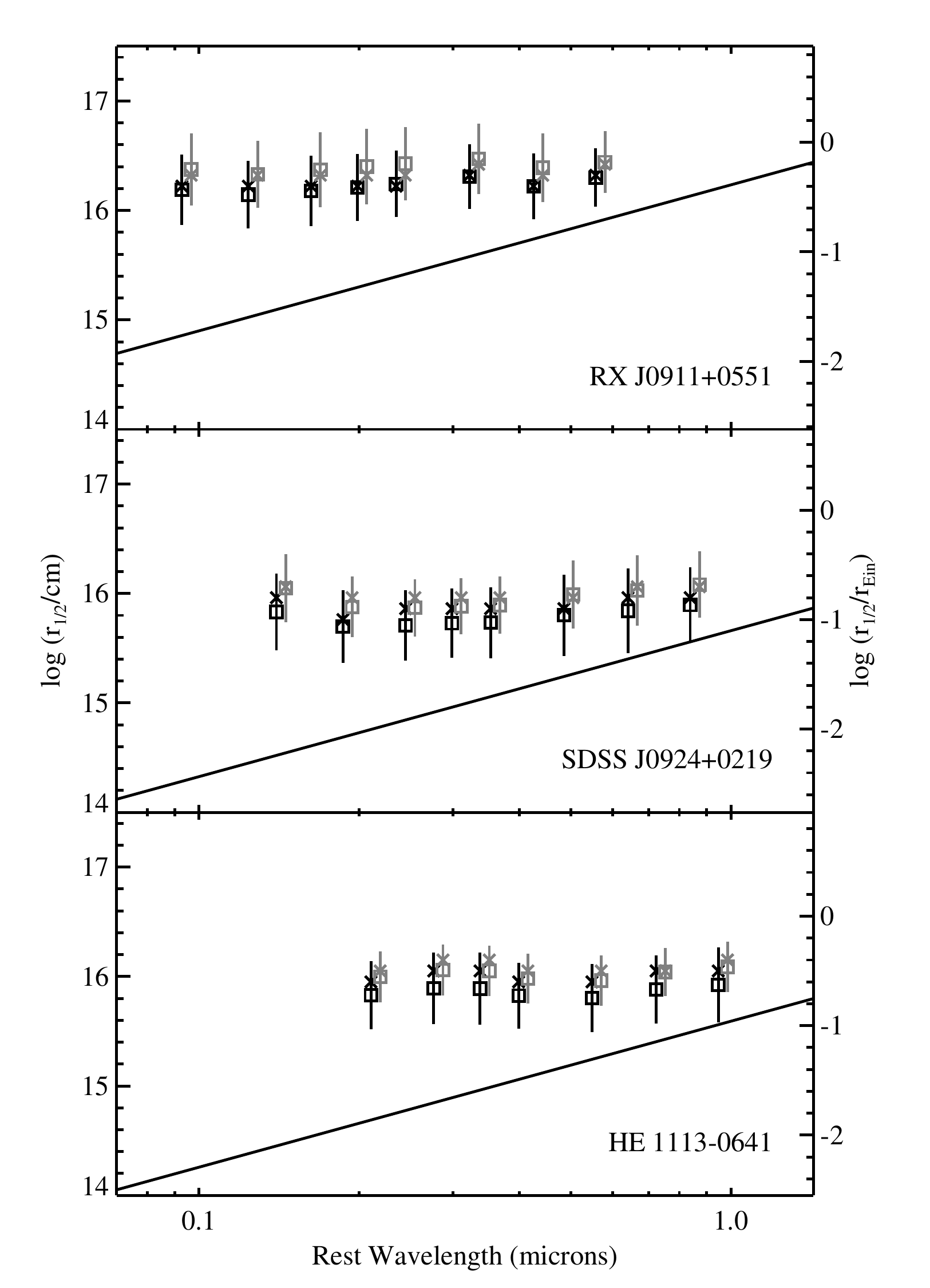}
  \caption{Medians (squares) and modes (crosses) of the probability
  distributions for the half-light source radius. The error bars
  correspond to the sum in quadratures of the intrinsic width of the
  distributions and the scatter from the Monte Carlo propagation of
  measurement uncertainties. Black (gray) points indicate a
  logarithmic (linear) prior. The solid line is the prediction for the
  half-light radius (in cm) of the standard thin disk model, for the
  estimated black hole mass.}
  \label{fig:size2}
\end{figure}

\begin{figure}
  \centering
  \includegraphics[width=0.45\textwidth]{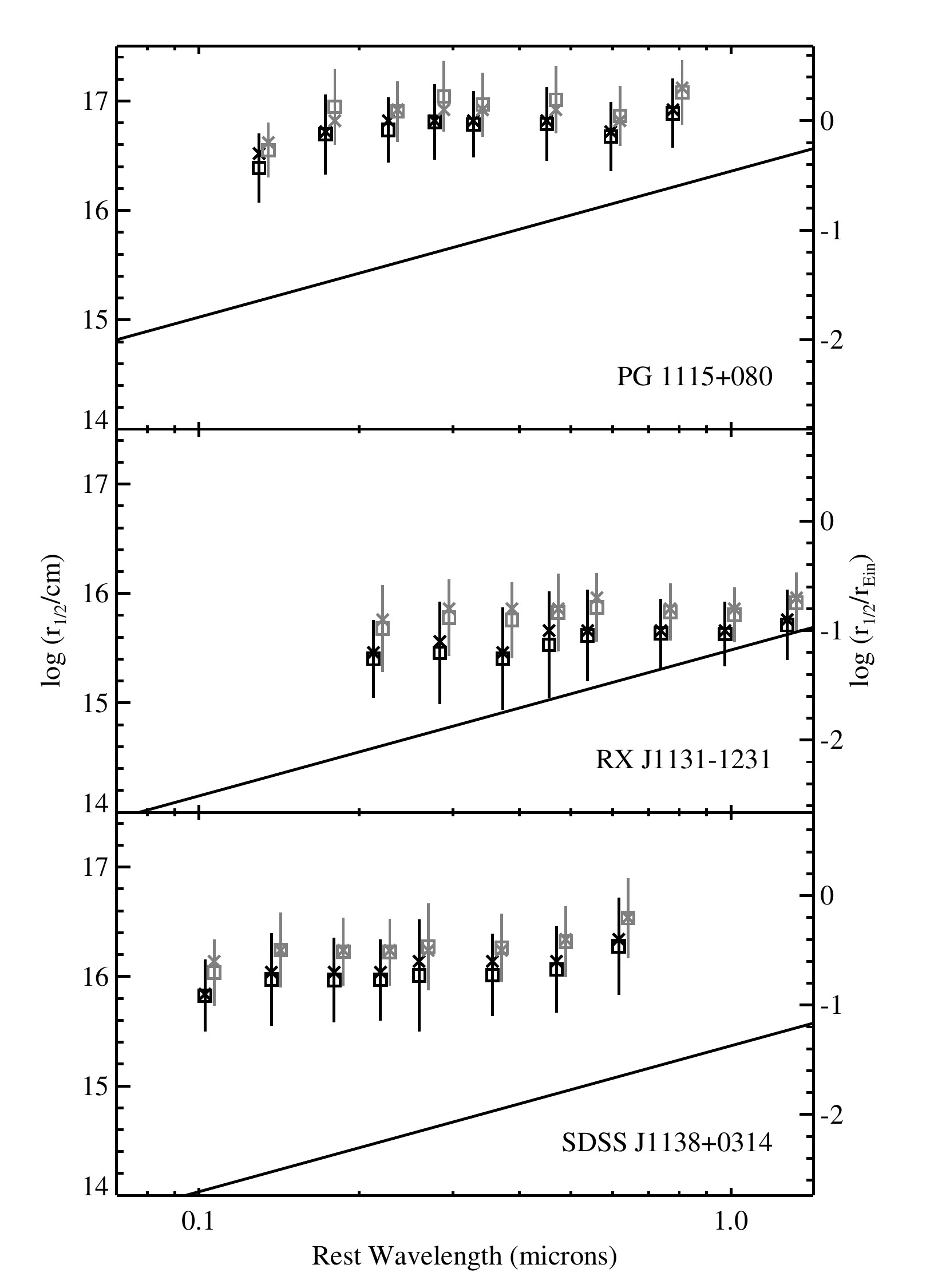}
  \caption{Medians (squares) and modes (crosses) of the probability
  distributions for the half-light source radius. The error bars
  correspond to the sum in quadratures of the intrinsic width of the
  distributions and the scatter from the Monte Carlo propagation of
  measurement uncertainties. Black (gray) points indicate a
  logarithmic (linear) prior. The solid line is the prediction for the
  half-light radius (in cm) of the standard thin disk model, for the
  estimated black hole mass.}
  \label{fig:size3}
\end{figure}

\begin{figure}
  \centering
  \includegraphics[width=0.45\textwidth]{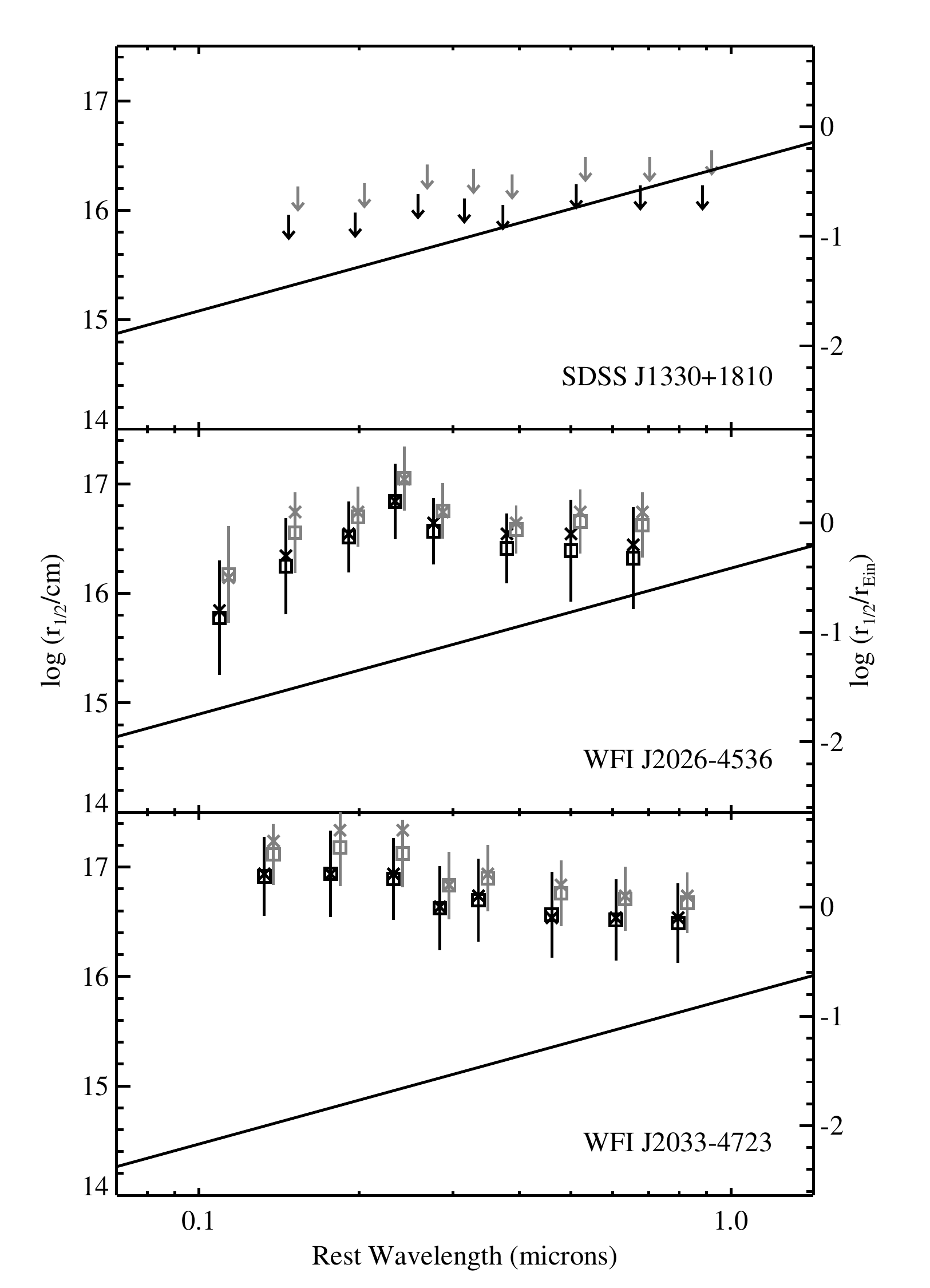}
  \caption{Medians (squares) and modes (crosses) of the probability
  distributions for the half-light source radius. The error bars
  correspond to the sum in quadratures of the intrinsic width of the
  distributions and the scatter from the Monte Carlo propagation of
  measurement uncertainties. Black (gray) points indicate a
  logarithmic (linear) prior. The solid line is the prediction for the
  half-light radius (in cm) of the standard thin disk model, for the
  estimated black hole mass. In the case of \sdssonethree, we only
  have upper limits on the size of the disk.}
  \label{fig:size4}
\end{figure}

The result of the analysis described in Section~\ref{sec:analysis} was
a posterior probability distribution for the half-light radius of the
quasar accretion disk in each of our 12 lensed quasars, in each
filter. The medians and modes of these distributions are listed, along
with $1\sigma$ confidence intervals, in Table~\ref{tab:sizes}, and
plotted in Figures~\ref{fig:size1}--\ref{fig:size4} as a function of
rest wavelength. The error bars are taken from the diagonal elements
of the covariance matrix $\mathcal{W}^{-1}$ defined in
Equation~(\ref{eqn:weightmatrix}). The choice of prior for the
half-light radii has a small effect on the measured values. In
general, the linear prior favors values of $r_{1/2} \sim 0.2$ dex
larger than the logarithmic prior does. This is to be expected, since
the linear prior has more probability density at larger sizes. The
slope with wavelength is generally unaffected by the choice of
prior. Since a logarithmic prior is more appropriate for a scale-free
quantity such as the half-light radius, we adopt these values.

The half-light radii are measured in terms of the Einstein radius of a
solar-mass microlens $r_\mathrm{Ein}$. We converted these angular
sizes to physical radii in the source plane using the angular diameter
distances $D_\mathrm{OS}$, $D_\mathrm{OL}$, and $D_\mathrm{LS}$
(observer-to-source, observer-to-lens, and lens-to-source,
respectively). The physical source radius depends weakly on the
redshifts of the source and lens, and the chosen cosmology. It also
depends on our choice of the stellar mass function, going as the
square root of the mean mass (see Section~\ref{sec:magmaps}). We show
physical distances on the left axis of
Figures~\ref{fig:size1}--\ref{fig:size4}, with the corresponding
fraction of an Einstein radius on the right axis.

In order to compare the measured source sizes to the thin disk
predictions, we must estimate the Eddington fraction, accretion
efficiency, and black hole masses of our sample of quasars. Following
\citet{Pooley:2007p19}, we set $f_\mathrm{Edd} = 0.25$
\citep{Kollmeier:2006p128} and $\eta = 0.15$ \citep{Yu:2002p965}. We
turned to the literature to find black hole mass estimates for most of
our sample. For \mgfour, \hefour, \rxnine, \sdssnine, \pgoneone, and
\rxoneone\ we adopted the virial mass estimates of
\citet{Peng:2006p616}, which are based on the widths of the broad
emission lines C\textsc{iv}, Mg\textsc{ii}, and H$\beta$. For
\sdssoneone\ we adopted that of \citet{Morgan:2010p1129}, calculated
using the same method. We adopted the masses estimated by
\citet{Pooley:2007p19} for \hetwo\ and \wfithree\ using their optical
and X-ray luminosities. For the remaining quasars (\heoneone,
\sdssonethree, and \wfitwosix) we used the luminosity method of
\citet{Pooley:2007p19}, inferring a bolometric luminosity from the
optical flux. The optical estimates used the relation
\begin{equation}
L_\mathrm{bol} = 9[\lambda f_\lambda]_{5100\text{\AA}} 4 \pi d_l^2
\end{equation}
\citep{Kaspi:2000p631}, where $d_l$ is the luminosity distance to the
quasar, and the rest-frame 5100\,\AA\ flux is extrapolated from the
magnification-corrected \textit{HST} NICMOS $F160W$ flux of the LM
image using an assumed power-law spectrum $f_\lambda \sim
\lambda^{-1.7}$ \citep{Kollmeier:2006p128}. We used our assumed
Eddington ratio $f_\mathrm{Edd} = 0.25$ to calculate black hole
masses. The luminosity and virial mass estimates for our sample of
quasars are listed in Table~\ref{tab:bhmass}. We adopted the virial
estimates where available, and the luminosity masses elsewhere. We
note that in cases where both are available, the luminosity-based
estimates are systematically smaller than the virial estimates by
$\sim 0.5$ dex; this could lead to under-predictions of the accretion
disk size by factors up to $\sim 2$ for the cases where we use the
luminosity estimate in the absence of a virial estimate. This
systematic bias is a little surprising, since the bolometric
luminosity technique is calibrated using the virial method
\citep{Kollmeier:2006p128}. It may be partially due to a bias in our
magnification correction, but it is difficult to imagine that we have
systematically overestimated magnifications by a factor of three.

Using these black hole mass estimates, we plot the prediction of the
thin disk model for each quasar in
Figures~\ref{fig:size1}--\ref{fig:size4}. Comparing the microlensing
sizes to these curves, it is immediately apparent that the measured
sizes are systematically larger than predicted by the thin disk model
and vary more weakly with wavelength than predicted.

\addtocounter{table}{1}
\begin{deluxetable*}{lcccc}
\tablewidth{0pt}
\tablecaption{Black Hole Mass Estimates
  \label{tab:bhmass}}
\tablehead{
  \colhead{} & 
  \colhead{$L_\mathrm{bol,opt}$} &
  \colhead{$L_\mathrm{bol,X}$} &
  \colhead{$M_\mathrm{BH,opt}$} &
  \colhead{$M_\mathrm{BH,vir}$} \\
  \colhead{Quasar} & 
  \colhead{($10^{46}$ erg s$^{-1}$)} &
  \colhead{($10^{46}$ erg s$^{-1}$)} &
  \colhead{($10^9 M_\odot$)} &
  \colhead{($10^9 M_\odot$)}
}
\startdata
\hetwo        & 0.29                             & 0.63                             & 0.092       & \nodata                            \\
\mgfour       & 3.6\phn                          & 2.8\phn                          & 1.1\phn\phn & 1.82                               \\
\hefour       & 0.38\makebox[0pt][l]{\tnm{a}}    & 0.46\makebox[0pt][l]{\tnm{a}}    & 0.12\phn    & 0.50                               \\
\rxnine       & 1.3\phn                          & 1.3\phn                          & 0.41\phn    & 0.80                               \\
\sdssnine     & 0.06                             & 0.03                             & 0.019       & 0.11                               \\
\heoneone     & 0.27\makebox[0pt][l]{\tnm{a}}    & 0.10\makebox[0pt][l]{\tnm{a}}    & 0.087       & \nodata                            \\
\pgoneone     & 1.1\phn                          & 0.66                             & 0.35\phn    & 0.92/1.23\makebox[0pt][l]{\tnm{b}} \\
\rxoneone     & 0.08                             & 0.13                             & 0.025       & 0.06                               \\
\sdssoneone   & 0.38\makebox[0pt][l]{\tnm{a}}    & 0.25\makebox[0pt][l]{\tnm{a}}    & 0.12\phn    & 0.04\makebox[0pt][l]{\tnm{c}}      \\
\sdssonethree & 4.7\makebox[0pt][l]{\tnm{a}}\phn & \nodata                          & 1.5\phn\phn & \nodata                            \\
\wfitwosix    & 2.5\makebox[0pt][l]{\tnm{a}}\phn & 1.1\makebox[0pt][l]{\tnm{a}}\phn & 0.79\phn    & \nodata                            \\
\wfithree     & 0.57                             & 0.38                             & 0.18\phn    & \nodata
\enddata
\tablenotetext{a}{This work.}
\tablenotetext{b}{Two values are from the C\,\textsc{iv} and
  Mg\,\textsc{ii} lines, respectively. We adopt the Mg\,\textsc{ii}
  value.}
\tablenotetext{c}{\protect{\citet{Morgan:2010p1129}}}
\tablecomments{Unless otherwise indicated,
  bolometric luminosity estimates are from
  \protect{\citet{Pooley:2007p19}} and virial mass estimates are
  from \protect{\citet{Peng:2006p616}}.}
\end{deluxetable*}

\begin{deluxetable*}{lcccccccc}
\tabletypesize{\small}
\tablewidth{0pt}
\tablecaption{Best-fit Parameters for Disk Size Versus Wavelength
  \label{tab:params}
}
\tablehead{
  \colhead{} &
  \colhead{} &
  \multicolumn{3}{c}{Logarithmic Prior} & 
  \colhead{} &
  \multicolumn{3}{c}{Linear Prior} \\
  \cline{3-5}
  \cline{7-9} \\[-9pt]
  \colhead{Quasar} & 
  \colhead{$\lambda_c$\tnm{a}(\AA)} &
  \colhead{$Y_c$} &
  \colhead{$\log(r_{1/2}/r_\mathrm{pred})$} &
  \colhead{$\nu$} &
  \colhead{} &
  \colhead{$Y_c$} &
  \colhead{$\log(r_{1/2}/r_\mathrm{pred})$} &
  \colhead{$\nu$}
}
\startdata
\hetwo      & 2763 & $16.57\pm0.15$ & $+1.71$ & $-0.56\pm0.47$ && $16.79\pm0.18$ & $+1.93$ & $-0.36\pm0.43$ \\
\mgfour     & 3075 & $15.90\pm0.19$ & $+0.11$ & $+1.50\pm0.84$ && $16.10\pm0.16$ & $+0.31$ & $+1.49\pm0.74$ \\
\hefour     & 3250 & $16.09\pm0.19$ & $+0.64$ & $+0.67\pm0.55$ && $16.37\pm0.16$ & $+0.93$ & $+0.55\pm0.49$ \\
\rxnine     & 2299 & $16.23\pm0.13$ & $+0.84$ & $+0.17\pm0.41$ && $16.39\pm0.13$ & $+1.01$ & $+0.12\pm0.42$ \\
\sdssnine   & 3462 & $15.79\pm0.16$ & $+0.75$ & $+0.17\pm0.49$ && $15.97\pm0.13$ & $+0.92$ & $+0.19\pm0.42$ \\
\heoneone   & 4438 & $15.86\pm0.18$ & $+0.74$ & $+0.05\pm0.49$ && $16.03\pm0.11$ & $+0.91$ & $+0.05\pm0.38$ \\
\pgoneone   & 3212 & $16.72\pm0.12$ & $+1.02$ & $+0.40\pm0.45$ && $16.90\pm0.11$ & $+1.20$ & $+0.45\pm0.39$ \\
\rxoneone   & 5270 & $15.55\pm0.14$ & $+0.43$ & $+0.40\pm0.50$ && $15.80\pm0.12$ & $+0.69$ & $+0.20\pm0.46$ \\
\sdssoneone & 2540 & $16.01\pm0.19$ & $+1.44$ & $+0.41\pm0.54$ && $16.26\pm0.16$ & $+1.68$ & $+0.43\pm0.45$ \\
\wfitwosix  & 2705 & $16.52\pm0.15$ & $+1.04$ & $+0.27\pm0.53$ && $16.68\pm0.12$ & $+1.20$ & $+0.17\pm0.42$ \\
\wfithree   & 3285 & $16.71\pm0.16$ & $+1.55$ & $-0.63\pm0.52$ && $16.91\pm0.13$ & $+1.75$ & $-0.67\pm0.41$ 
\enddata
\tablenotetext{a}{$\lambda_c$ is the geometric average of the rest
  wavelengths of our observations.}
\end{deluxetable*}

For each lens we fit a power-law model $Y(\lambda)$ to the medians of
the distributions:
\begin{equation}
\label{eqn:fit}
Y(\lambda) = Y_c
+\nu\log\left(\frac{\lambda}{\lambda_c}\right) ~,
\end{equation}
where the central wavelength $\lambda_c$ is the geometric mean of
the rest wavelengths and $Y_c$ is the logarithm of the half-light
radius of the source at that wavelength, in cm. Since the errors on the
medians are correlated, we minimized the statistic
\begin{equation}
\label{eqn:chisq}
\chi^2 = \left(\mathbf{y} - Y(\boldsymbol{\lambda})\right)^\mathrm{T} \cdot \mathcal{W} \cdot
\left(\mathbf{y} - Y(\boldsymbol{\lambda})\right) ~,
\end{equation}
where $\mathbf{y}$ is the vector of median sizes measured at
wavelengths $\boldsymbol\lambda$, and $\mathcal{W}$ is the weight
matrix defined in Equation~(\ref{eqn:weightmatrix}). The parameters
$Y_c$ and $\nu$ of the best-fit power laws are listed in
Table~\ref{tab:params}.

In nearly every case, the measured disk size at the central wavelength
$\lambda_c$ is larger than the thin disk model predicts, by factors
ranging from $\sim 2$ to more than 30. The logarithm of this factor is
listed for each quasar in Table~\ref{tab:params}. Comparing the
logarithmic offset to our errors in $Y_c$, we have ruled out the thin
disk normalization by at least $3\sigma$ in all cases except \mgfour,
and as many as $10\sigma$ in some cases. This result is consistent
with that of \citet{Pooley:2007p19}, though our analysis method is
more quantitative. The average offset between $Y_c$ and the thin disk
prediction at the same wavelength is 0.89 dex, or a factor of 7.5. A
chi-square test comparing the measured sizes to predicted sizes yields
$\chi^2/N_\mathrm{dof}=46$.

\begin{figure}
  \centering
  \includegraphics[width=0.45\textwidth]{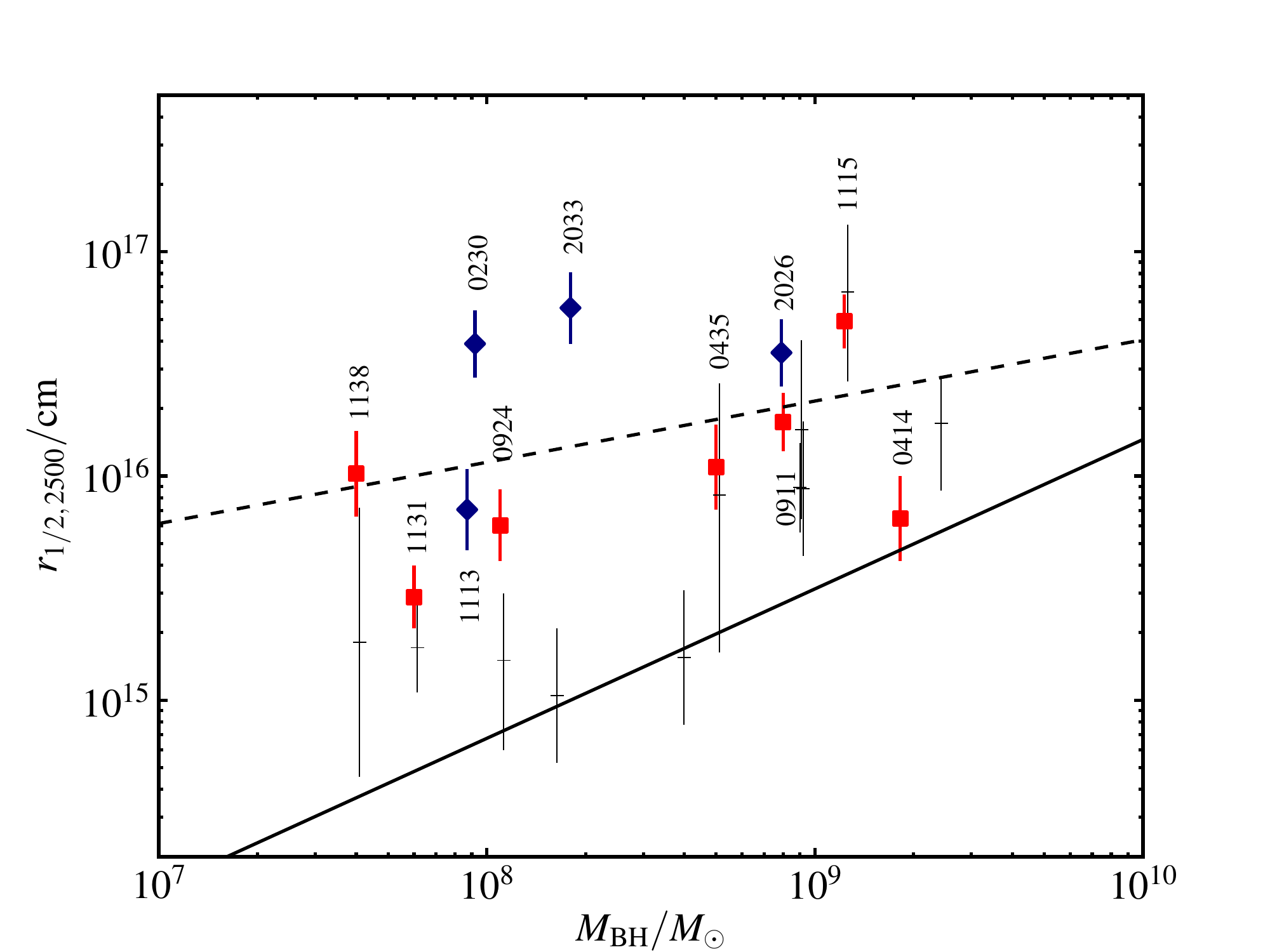}
  \caption{Half-light radii at a rest wavelength of 2500\,\AA, plotted
    against black hole mass. Black hole masses estimated using the
    virial method are plotted as red squares; those estimated using
    bolometric luminosity are plotted as blue diamonds. The solid line
    is the prediction of the thin disk model (with $f_\mathrm{Edd} =
    0.25$ and $\eta = 0.15$), while the dotted line is the best fit to
    these data. The microlensing radii of \citet{Morgan:2010p1129} are
    plotted in thin black lines for comparison. They have been
    corrected to 2500\,\AA\ (assuming that $r \propto \lambda^{4/3}$)
    and converted to half-light radii by multiplying by 2.44 (see
    Equation~(\protect{\ref{eqn:rhalf}})). The inclination correction
    has been removed. Their error bars are larger because they take
    several sources of systematic error into account. The black hole
    masses have unplotted uncertainties of about 0.4 dex.}
  \label{fig:sizemassplot}
\end{figure}

In Figure~\ref{fig:sizemassplot}, we plot the half-light disk radii of
our sample (excluding \sdssonethree) at fixed rest wavelength versus
the black hole mass. The disk radii are calculated using the best-fit
model (see Equation~(\ref{eqn:fit})) at $\lambda=2500$\,\AA. This
wavelength is chosen to match the one used by
\citet{Morgan:2010p1129}, who extrapolate from single-wavelength
microlensing observations to create a similar plot. The errors on the
half-light radii are the uncertainties on $Y_c$ (see
Table~\ref{tab:params}), and uncertainties of $0.4$ dex have been
assigned to the black hole masses (but not plotted). We multiplied the
mass uncertainties by the thin disk model slope of $2/3$ and added
them in quadrature to the uncertainties on the half-light radii. With
these uncertainties, the best-fit slope for $r_{1/2}$ versus
$M_\mathrm{BH}$ is $0.27\pm0.17$. The $\chi^2$ for this fit is $17.3$
for $9$ degrees of freedom. Thus, we find a slope that is about half
the expected $r_{1/2} \propto M^{2/3}_{\rm BH}$ dependence (see
Equation (\ref{eqn:rhalf})), but with a substantial
uncertainty. Perhaps more importantly, we find a significant offset
from the thin disk model in the overall normalization, nearly an order
of magnitude. The only quasar that we find to be consistent with the
prediction is \mgfour.

For comparison, we have plotted the microlensing size estimates of
\citet{Morgan:2010p1129} in Figure~\ref{fig:sizemassplot}, including
their wavelength corrections but undoing their inclination corrections
to match our convention and multiplying by a factor of $2.44$ to
convert to half-light radii (see Equation~(\ref{eqn:rhalf})). They
find a logarithmic offset between their microlensing sizes and the
thin disk prediction in the same sense as ours, and about half as
large. Though we do not have a good explanation for the offset between
our microlensing size measurements and the prediction, the smaller
offset between our sizes and those of \citet{Morgan:2010p1129} may
arise from the difference in analysis techniques; analyzing this
possibility is the subject of future work.

Increasing the Eddington fraction of the quasars, or decreasing their
accretion efficiencies, would help to reconcile our microlensing sizes
with the predictions of the thin disk model, though the adjustments
would need to be drastic, because of the weak dependence of the size
on these values. Increasing the black hole mass estimates
systematically by factors of $\sim 20$ or decreasing the mean stellar
mass in the lensing galaxies by factors of $\sim 50$ would also bring
about agreement, but systematic offsets at these levels seem unlikely.

With one exception, every quasar in our sample displays a power-law
slope $\nu$ flatter than the expected $4/3$ by $1.2$ to $2.6\sigma$ or
more (see Table~\ref{tab:params}). Although these do not individually
rule out the thin disk slope at high significance, the $10$ to $1$
preponderance of quasars with too-shallow slopes makes a convincing
case against it. Treating the measurement for each quasar as an
independent constraint yields an average value $\nu =
0.17\pm0.15$. Comparing the eleven measured slopes to $4/3$ yields a
$\chi^2/N_\mathrm{dof} = 6.1$. Since a multi-temperature blackbody
accretion disk with a power-law temperature profile has
$T_\mathrm{eff} \propto r^{-1/\nu}$, applying our results to this
model implies that the effective temperature is a steeply falling
function of radius. It is worth noting that nearly all the measured
slopes are consistent with $\nu = 0$, i.e., a source with a
wavelength-independent size, though the existence of such an object
seems highly unlikely, from both theoretical (e.g., energy
conservation) and observational (e.g., quasar variability)
standpoints.

\mgfour, the only quasar that matches the thin disk size prediction
(and the only radio-loud quasar in our sample), is also the only
quasar with a slope $\nu$ consistent with $4/3$. Although \mgfour\ was
the only case where we used empirical mid-IR flux ratios in place of
model ratios (as we suspected the presence of millilensing), it is
unlikely that its unique behavior is due to this choice, since mid-IR
radios for other lensed quasars (including \pgoneone) match the models
very well \citep{Chiba:2005p53,Minezaki:2009p610}.

Two lenses, \hetwo\ and \wfithree, have microlensing size estimates
that decrease with wavelength. This result does not seem to arise from
a gross failure of our analysis method, since the flux ratios
themselves become more anomalous with wavelength, which in most cases
implies a smaller source size (see Table~\ref{tab:optdata}). It cannot
be due to microlensing variability, since the optical/IR
measurements are coeval. But it should not be concluded that the
temperature profile of these quasars is in fact inverted, as the
errors do not rule out positive slopes. It is possible that the
sources lie on an unusual region of the caustic pattern where the
smoothing caused by a larger source causes {\em more} anomalous flux
ratios. These lenses underscore the importance of having a large
sample of objects when using single-epoch measurements.


\section{Summary and Conclusions}
\label{sec:conclusions}

We have obtained flux ratios in eight optical and IR filters spanning
a factor of six in wavelength, for a sample of 12 strongly lensed
quasars. Comparing these ratios to each other and to \chandra\ X-ray
flux ratios, both from the literature and newly reported here, we see
chromatic variations, which we attribute to the microlensing of a
multicolor accretion disk. The standard thin accretion disk model
predicts a temperature profile that falls as the $\beta = 3/4$ power
of radius, implying that the half-light radius of the disk is
proportional to $\lambda^{4/3}$, with an overall normalization that
depends on the black hole mass and (weakly) on the Eddington fraction
and accretion efficiency. This chromatic dependence combines with the
dependence of microlensing magnification on source size to produce
higher-amplitude microlensing variations at blue wavelengths than at
red wavelengths. For single-epoch observations like ours, this usually
means that the observed flux ratios at blue wavelengths are more
anomalous than at red wavelengths, compared to the ratios predicted by
smooth lens models.

In addition to formal statistical uncertainties, we have estimated the
systematic errors in the flux ratios due to spectral contamination by
broad emission lines, quasar variability combined with lensing time
delays, and microlensing variability combined with delays between
measurement epochs. We have described a Bayesian method for
determining the probability distribution of the half-light radius of
the source in each filter. For quasars without X-ray measurements, we
are able to place upper limits on the half-light radii, by virtue of
the departures of the flux ratios from the model predictions. When
X-ray ratios are available, we assume that they originate from a very
compact region, and we are able to place both upper and lower limits
on the half-light radius, reasoning that the differences between the
X-ray and optical ratios arise because of the appreciable optical
radius.

The resulting size estimates are larger overall than is predicted by
the standard thin disk model by nearly an order of magnitude, on
average. The scaling of the half-light radius with mass is consistent
with the expected slope, but the scaling with wavelength is shallower
than expected. Although the large error bars on the wavelength slopes
mean that a single slope measurement would have little weight, the
fact that all but one of the measured slopes are too shallow commands
attention. Since the scaling of radius with wavelength is the
reciprocal of the scaling of temperature with radius, our results
indicate that if we assume a blackbody accretion disk with a power-law
temperature profile the temperature slope is in general \emph{steeper}
than the standard thin disk model predicts. However, other
explanations for the relative insensitivity of radius to wavelength
are also possible, though none immediately present themselves.

There is growing evidence from microlensing studies that quasar
accretion disks are larger at optical wavelengths than simple
accretion models predict. Our result is a confirmation of earlier,
more qualitative work by \citet{Pooley:2007p19}. It also corroborates
the results of \citet{Morgan:2010p1129}, who use several lensed
quasars to study the mass dependence of the accretion disk radius, and
find it difficult to reconcile their radii with thin disk theory. Our
results for the observer-frame $r'$-band size of \mgfour\ and
$u'$-band size of \sdssnine\ are consistent with the upper limits of
\citet{Bate:2008p1955} and \citet{Floyd:2009p233}, respectively.
Likewise, our estimates of the radius of \rxoneone\ at around
4000\,\AA\ in the rest frame, and of \hefour\ at around 6500\,\AA\ in
the observed frame, are consistent with those of \citet{Dai:2010p278}
and \citet{Mosquera:2010p0}, respectively.

The picture is less clear regarding the temperature profile slope
$\beta$. Using time-domain measurements at several wavelengths,
\citet{Poindexter:2008p34} find $\beta \sim 0.6$ for the quasar
\mbox{HE\,$1104-1805$}, consistent with the thin disk slope of
$3/4$. They point out that a flatter temperature slope (i.e., $\beta <
3/4$) would help to explain the faintness of lensed quasars relative
to the expected flux from a Shakura--Sunyaev disk of the size implied
by microlensing. However, our result is just the opposite: the low
average value of $\nu$ implies that the temperature slope $\beta$ is
greater than $3/4$. Other single-lens microlensing studies have
addressed this question as well. \citet{Bate:2008p1955} and
\citet{Floyd:2009p233} use single-epoch broadband photometry similar
to ours to constrain the slope $\nu$ for the quasars \mgfour\ and
\sdssnine, respectively; their limits are compatible both with our
results and with the thin disk model. \citet{Mosquera:2010p0} use
narrowband imaging of \hefour, finding a slope $\nu = 1.3 \pm 0.3$,
consistent with our slope within the errors. In addition, several
studies have focused on \qtwotwo, because of the short time scale and
large amplitude of its microlensing variability, and the large amount
of monitoring data that are available. \citet{Anguita:2008p327} find a
ratio of $r'$- to $g'$-band source sizes consistent with the thin disk
model. \citet{Eigenbrod:2008p933} use spectrophotometric monitoring to
constrain $\nu = 1.2 \pm 0.3$, and \citet{Mosquera:2009p1292} use
narrowband photometry, finding solutions compatible with $\nu = 4/3$,
as well as with higher values. Our result, working with a larger
sample size, is the first to indicate that $\nu$ is significantly less
than $4/3$, so further study is clearly needed. We note that the error
bars on $\nu$ are uniformly large for individual lenses. Increasing
the sample size, as we have done, is a good way to improve this;
additional strategies include using spectroscopic methods to transform
line emission from a nuisance into a tool, improving the coordination
between X-ray and optical observing campaigns, and moving to
multiwavelength, long temporal baseline monitoring.

\acknowledgements

The authors thank Joachim Wambsganss for the use of his microlensing
simulation code. J.A.B. and P.L.S. acknowledge the support of US NSF
grant AST-0607601.


\bibliography{ms}


\clearpage
\setcounter{table}{4}
\LongTables
\begin{deluxetable*}{cccccc}
\tablewidth{0pt}
\tablecaption{Relative Photometry
  \label{tab:optdata}}
\tablehead{
  \colhead{Quasar/Filter} & 
  \colhead{HM\tnm{a}} &
  \colhead{HS} & 
  \colhead{LM} & 
  \colhead{LS} &
  \colhead{rms\tnm{b}}
}
\startdata
\multicolumn{1}{l}{\hetwolong}
          & A                     & B                     & C                     & D                     & \\ \cline{1-6}
$u'$      & $0\pm0.00$            & $+0.26\pm0.03\pm0.00$ & $+0.97\pm0.03\pm0.01$ & $+2.60\pm0.08\pm0.03$ & 0.17 \\
$g'$      & $0\pm0.00$            & $+0.25\pm0.01\pm0.05$ & $+0.69\pm0.01\pm0.05$ & $+2.41\pm0.04\pm0.11$ & 0.14 \\
$r'$      & $0\pm0.00$            & $+0.23\pm0.01\pm0.05$ & $+0.62\pm0.01\pm0.05$ & $+2.44\pm0.02\pm0.20$ & 0.17 \\
$i'$      & $0\pm0.00$            & $+0.17\pm0.01\pm0.00$ & $+0.58\pm0.01\pm0.01$ & $+2.37\pm0.05\pm0.36$ & 0.15 \\
$z'$      & $0\pm0.00$            & $+0.12\pm0.01\pm0.05$ & $+0.55\pm0.02\pm0.05$ & $+2.95\pm0.15\pm0.57$ & 0.38 \\
$J$       & $0\pm0.00$            & $+0.15\pm0.01\pm0.00$ & $+0.54\pm0.01\pm0.01$ & $+2.64\pm0.07\pm0.94$ & 0.26 \\
$H$       & $0\pm0.00$            & $+0.12\pm0.01\pm0.05$ & $+0.50\pm0.01\pm0.05$ & $+2.74\pm0.12\pm1.18$ & 0.30 \\
$K_s$     & $0\pm0.00$            & $+0.07\pm0.01\pm0.05$ & $+0.50\pm0.01\pm0.05$ & $+3.48\pm0.16\pm1.62$ & 0.62 \\
0.5--8keV (ObsID 1642) & $-0.51\pm0.13\pm0.50$ & $+0.39\pm0.19\pm1.20$ & $0\pm0.30$ & $+0.87\pm0.18\pm0.50$ & 0.60 \\
\cline{1-6}\\[-6pt]
\multicolumn{1}{l}{\mgfourlong}
          & A1                    & A2                    & B                     & C                     & \\ \cline{1-6}
$r'$      & $+0.27\pm0.22\pm0.05$ & $+0.21\pm0.21\pm0.05$ & $0\pm0.02$            & $+0.43\pm0.11\pm0.07$ & 0.88 \\
$i'$      & $-0.50\pm0.09\pm0.00$ & $+0.30\pm0.16\pm0.00$ & $0\pm0.02$            & $+1.03\pm0.10\pm0.04$ & 0.57 \\
$z'$      & $-0.66\pm0.03\pm0.00$ & $+0.02\pm0.06\pm0.00$ & $0\pm0.02$            & $+1.02\pm0.05\pm0.04$ & 0.46 \\
$J$       & $-0.93\pm0.01\pm0.00$ & $-0.41\pm0.01\pm0.00$ & $0\pm0.02$            & $+0.94\pm0.01\pm0.04$ & 0.31 \\
$H$       & $-0.96\pm0.01\pm0.05$ & $-0.72\pm0.01\pm0.05$ & $0\pm0.02$            & $+0.90\pm0.01\pm0.07$ & 0.21 \\
$K_s$     & $-1.01\pm0.01\pm0.00$ & $-0.86\pm0.01\pm0.00$ & $0\pm0.02$            & $+0.93\pm0.01\pm0.04$ & 0.15 \\
0.5--8keV (ObsID 3919) & $-0.81\pm0.05\pm0.50$ & $-0.29\pm0.08\pm1.20$ & $0\pm0.35$ & $+0.94\pm0.05\pm0.40$ & 0.36 \\
\cline{1-6}\\[-6pt]
\multicolumn{1}{l}{\hefourlong}
          & C                     & B                     & A                     & D                     & \\ \cline{1-6}
$u'$      & $+0.93\pm0.02\pm0.01$ & $+1.01\pm0.02\pm0.01$ & $0\pm0.00$            & $+0.90\pm0.02\pm0.02$ & 0.45 \\
$g'$      & $+0.73\pm0.01\pm0.05$ & $+0.77\pm0.01\pm0.05$ & $0\pm0.00$            & $+0.85\pm0.01\pm0.05$ & 0.35 \\
$r'$      & $+0.69\pm0.02\pm0.01$ & $+0.70\pm0.02\pm0.01$ & $0\pm0.00$            & $+0.83\pm0.03\pm0.02$ & 0.32 \\
$i'$      & $+0.56\pm0.01\pm0.05$ & $+0.62\pm0.01\pm0.05$ & $0\pm0.00$            & $+0.84\pm0.01\pm0.05$ & 0.28 \\
$z'$      & $+0.54\pm0.02\pm0.01$ & $+0.58\pm0.02\pm0.01$ & $0\pm0.00$            & $+0.78\pm0.02\pm0.02$ & 0.27 \\
$J$       & $+0.49\pm0.01\pm0.05$ & $+0.50\pm0.01\pm0.05$ & $0\pm0.00$            & $+0.69\pm0.01\pm0.05$ & 0.24 \\
$H$       & $+0.44\pm0.01\pm0.05$ & $+0.47\pm0.01\pm0.05$ & $0\pm0.00$            & $+0.67\pm0.01\pm0.05$ & 0.23 \\
$K_s$     & $+0.36\pm0.01\pm0.01$ & $+0.38\pm0.01\pm0.01$ & $0\pm0.00$            & $+0.64\pm0.01\pm0.02$ & 0.19 \\
0.5--8keV (ObsID 7761) & $+1.06\pm0.13$ & $+1.07\pm0.14\pm0.35$ & $0\pm0.15$ & $+1.10\pm0.13\pm0.20$ & 0.47 \\
\cline{1-6}\\[-6pt]
\multicolumn{1}{l}{\rxninelong}
          & B                     & A                     & D                     & C                     & \\ \cline{1-6}
$u'$      & $+0.37\pm0.02\pm0.00$ & $0\pm0.00$            & $+1.64\pm0.01\pm0.09$ & $+1.31\pm0.02\pm0.00$ & 0.46 \\
$g'$      & $+0.26\pm0.01\pm0.10$ & $0\pm0.00$            & $+1.40\pm0.01\pm0.13$ & $+1.18\pm0.01\pm0.10$ & 0.43 \\
$r'$      & $+0.28\pm0.01\pm0.05$ & $0\pm0.00$            & $+1.33\pm0.01\pm0.10$ & $+1.14\pm0.01\pm0.05$ & 0.44 \\
$i'$      & $+0.13\pm0.01\pm0.05$ & $0\pm0.00$            & $+1.20\pm0.01\pm0.10$ & $+0.95\pm0.01\pm0.05$ & 0.39 \\
$z'$      & $+0.24\pm0.01\pm0.00$ & $0\pm0.00$            & $+1.22\pm0.01\pm0.09$ & $+1.00\pm0.01\pm0.00$ & 0.42 \\
$J$       & $+0.06\pm0.02\pm0.00$ & $0\pm0.00$            & $+1.09\pm0.01\pm0.09$ & $+0.79\pm0.01\pm0.00$ & 0.36 \\
$H$       & $+0.09\pm0.02\pm0.00$ & $0\pm0.00$            & $+1.02\pm0.02\pm0.09$ & $+0.77\pm0.02\pm0.00$ & 0.39 \\
$K_s$     & $-0.02\pm0.02\pm0.00$ & $0\pm0.00$            & $+0.94\pm0.03\pm0.09$ & $+0.77\pm0.02\pm0.00$ & 0.38 \\
0.5--8keV (ObsID 419) & $-0.29\pm0.03\pm0.45$ & $-1.33\pm0.17\pm0.80$ & $0\pm0.31$ & $+1.14\pm0.37\pm0.65$ & 0.98 \\
\cline{1-6}\\[-6pt]
\multicolumn{1}{l}{\sdssninelong}
          & A                     & D                     & B                     & C                     & \\ \cline{1-6}
$u'$      & $0\pm0.00$            & $+1.66\pm0.06\pm0.50$ & $+1.48\pm0.03\pm0.05$ & $+2.66\pm0.09\pm0.50$ & 0.71 \\
$g'$      & $0\pm0.00$            & $+2.76\pm0.03\pm0.05$ & $+1.53\pm0.01\pm0.05$ & $+2.59\pm0.02\pm0.05$ & 1.00 \\
$r'$      & $0\pm0.00$            & $+2.96\pm0.03\pm0.00$ & $+1.43\pm0.01\pm0.01$ & $+2.42\pm0.01\pm0.01$ & 1.07 \\
$i'$      & $0\pm0.00$            & $+2.96\pm0.03\pm0.05$ & $+1.36\pm0.01\pm0.05$ & $+2.28\pm0.01\pm0.05$ & 1.07 \\
$z'$      & $0\pm0.00$            & $+2.85\pm0.04\pm0.00$ & $+1.36\pm0.01\pm0.01$ & $+2.28\pm0.02\pm0.01$ & 1.03 \\
$J$       & $0\pm0.00$            & $+2.23\pm0.06\pm0.05$ & $+1.16\pm0.02\pm0.05$ & $+1.88\pm0.04\pm0.05$ & 0.81 \\
$H$       & $0\pm0.00$            & $+2.11\pm0.07\pm0.05$ & $+1.13\pm0.03\pm0.05$ & $+1.65\pm0.04\pm0.05$ & 0.75 \\
$K_s$     & $0\pm0.00$            & $+1.88\pm0.06\pm0.00$ & $+1.03\pm0.03\pm0.01$ & $+1.41\pm0.04\pm0.01$ & 0.68 \\
0.5--8keV (ObsID 5604) & $-1.26\pm0.23\pm0.30$ & $+0.87\pm0.54\pm0.70$ & $0\pm0.25$ & $+0.94\pm0.42\pm0.50$ & 0.78 \\
\cline{1-6}\\[-6pt]
\multicolumn{1}{l}{\heoneonelong}
          & B                     & D                     & A                     & C                     & \\ \cline{1-6}
$g'$      & $+0.10\pm0.01\pm0.05$ & $+0.81\pm0.01\pm0.05$ & $0\pm0.00$            & $+0.43\pm0.01\pm0.05$ & 0.43 \\
$r'$      & $+0.09\pm0.01\pm0.05$ & $+0.82\pm0.01\pm0.05$ & $0\pm0.00$            & $+0.50\pm0.01\pm0.05$ & 0.43 \\
$i'$      & $+0.07\pm0.01\pm0.00$ & $+0.83\pm0.01\pm0.00$ & $0\pm0.00$            & $+0.45\pm0.01\pm0.00$ & 0.44 \\
$z'$      & $+0.06\pm0.01\pm0.00$ & $+0.86\pm0.01\pm0.00$ & $0\pm0.00$            & $+0.41\pm0.01\pm0.00$ & 0.46 \\
$J$       & $+0.12\pm0.01\pm0.00$ & $+0.75\pm0.01\pm0.00$ & $0\pm0.00$            & $+0.39\pm0.01\pm0.00$ & 0.41 \\
$H$       & $+0.08\pm0.01\pm0.00$ & $+0.60\pm0.01\pm0.00$ & $0\pm0.00$            & $+0.42\pm0.01\pm0.00$ & 0.35 \\
$K_s$     & $+0.11\pm0.01\pm0.00$ & $+0.48\pm0.01\pm0.00$ & $0\pm0.00$            & $+0.55\pm0.01\pm0.00$ & 0.29 \\
0.5--8keV (ObsID 7760) & $+0.50\pm0.84\pm0.00$ & $+0.27\pm0.47\pm0.00$ & $0\pm0.00$ & $+1.75\pm1.19\pm0.01$ & 0.52 \\ 
\pagebreak
\cline{1-6}\\[-6pt]
\multicolumn{1}{l}{\pgoneonelong}
          & A1                    & A2                    & C                     & B                     & \\ \cline{1-6}
$u'$      & $-1.46\pm0.01\pm0.05$ & $-0.85\pm0.01\pm0.05$ & $0\pm0.02$            & $+0.40\pm0.01\pm0.05$ & 0.25 \\
$g'$      & $-1.45\pm0.01\pm0.05$ & $-1.28\pm0.01\pm0.05$ & $0\pm0.02$            & $+0.31\pm0.01\pm0.05$ & 0.10 \\
$r'$      & $-1.41\pm0.01\pm0.00$ & $-1.23\pm0.01\pm0.00$ & $0\pm0.02$            & $+0.35\pm0.01\pm0.02$ & 0.11 \\
$i'$      & $-1.42\pm0.01\pm0.05$ & $-1.22\pm0.01\pm0.05$ & $0\pm0.02$            & $+0.39\pm0.01\pm0.05$ & 0.10 \\
$z'$      & $-1.39\pm0.01\pm0.00$ & $-1.21\pm0.01\pm0.00$ & $0\pm0.02$            & $+0.38\pm0.01\pm0.02$ & 0.10 \\
$J$       & $-1.43\pm0.01\pm0.05$ & $-1.18\pm0.01\pm0.05$ & $0\pm0.02$            & $+0.41\pm0.01\pm0.05$ & 0.11 \\
$H$       & $-1.43\pm0.02\pm0.05$ & $-1.17\pm0.02\pm0.05$ & $0\pm0.02$            & $+0.34\pm0.02\pm0.05$ & 0.13 \\
$K_s$     & $-1.40\pm0.03\pm0.00$ & $-1.16\pm0.03\pm0.00$ & $0\pm0.02$            & $+0.48\pm0.05\pm0.02$ & 0.10 \\
0.5--8keV (ObsID 7757) & $-2.54\pm0.07\pm0.00$ & $-2.25\pm0.09\pm0.00$ & $-1.26\pm0.06\pm0.02$ & $-1.30\pm0.06\pm0.02$ & 0.34 \\
\cline{1-6}\\[-6pt]
\multicolumn{1}{l}{\rxoneonelong}
          & B                     & A                     & C                     & D                     & \\ \cline{1-6}
$u'$      & $-1.29\pm0.01\pm0.00$ & $-1.74\pm0.01\pm0.02$ & $0\pm0.01$            & $+0.63\pm0.02\pm0.07$ & 0.74 \\
$g'$      & $-1.06\pm0.01\pm0.05$ & $-1.52\pm0.01\pm0.05$ & $0\pm0.01$            & $+0.94\pm0.01\pm0.08$ & 0.62 \\
$r'$      & $-1.10\pm0.04$        & $-1.55\pm0.06$        & $0\pm0.01$            & $+1.00\pm0.11$        & 0.54 \\
$i'$      & $-1.00\pm0.10$        & $-1.46\pm0.12$        & $0\pm0.01$            & $+0.97\pm0.10$        & 0.58 \\
$z'$      & $-0.90\pm0.08$        & $-1.33\pm0.11$        & $0\pm0.01$            & $+1.00\pm0.08$        & 0.60 \\
$J$       & $-0.78\pm0.17$        & $-1.21\pm0.23$        & $0\pm0.01$            & $+0.75\pm0.12$        & 0.74 \\
$H$       & $-0.64\pm0.18$        & $-1.04\pm0.22$        & $0\pm0.01$            & $+0.65\pm0.14$        & 0.83 \\
$K_s$     & $-0.51\pm0.12$        & $-1.12\pm0.20$        & $0\pm0.01$            & $+0.97\pm0.18$        & 0.77 \\
0.5--8keV (ObsID 7787) & $-1.26\pm0.08\pm0.00$ & $-1.89\pm0.08\pm0.02$ & $0\pm0.01$ & $+0.51\pm0.11\pm0.07$ & 0.78 \\
0.5--8keV (ObsID 7789) & $-1.16\pm0.08\pm0.00$ & $-1.87\pm0.07\pm0.02$ & $0\pm0.01$ & $+1.07\pm0.12\pm0.07$ & 0.63 \\
\cline{1-6}\\[-6pt]
\multicolumn{1}{l}{\sdssoneonelong}
          & A                     & D                     & C                     & B                     & \\ \cline{1-6}
$u'$      & $0\pm0.01$            & $+1.98\pm0.05\pm0.05$ & $+1.70\pm0.03\pm0.05$ & $+1.35\pm0.03\pm0.05$ & 0.72 \\
$g'$      & $0\pm0.01$            & $+1.16\pm0.01\pm0.10$ & $+1.28\pm0.01\pm0.10$ & $+1.33\pm0.01\pm0.10$ & 0.42 \\
$r'$      & $0\pm0.01$            & $+1.28\pm0.01\pm0.05$ & $+1.27\pm0.01\pm0.05$ & $+1.38\pm0.01\pm0.05$ & 0.45 \\
$i'$      & $0\pm0.01$            & $+1.36\pm0.01\pm0.01$ & $+1.28\pm0.01\pm0.00$ & $+1.37\pm0.01\pm0.01$ & 0.47 \\
$z'$      & $0\pm0.01$            & $+1.17\pm0.01\pm0.05$ & $+1.19\pm0.01\pm0.05$ & $+1.21\pm0.01\pm0.05$ & 0.41 \\
$J$       & $0\pm0.01$            & $+1.33\pm0.02\pm0.01$ & $+1.16\pm0.01\pm0.00$ & $+1.39\pm0.02\pm0.01$ & 0.45 \\
$H$       & $0\pm0.01$            & $+1.26\pm0.03\pm0.05$ & $+1.06\pm0.02\pm0.05$ & $+1.33\pm0.03\pm0.05$ & 0.42 \\
$K_s$     & $0\pm0.01$            & $+0.83\pm0.03\pm0.05$ & $+0.89\pm0.03\pm0.05$ & $+1.23\pm0.04\pm0.05$ & 0.27 \\
0.5--8keV (ObsID 7759) & $-1.26\pm0.35\pm0.01$ & $-0.29\pm0.47\pm0.01$ & $0\pm0.00$ & $+0.00\pm0.46\pm0.01$ & 0.37 \\
\cline{1-6}\\[-6pt]
\multicolumn{1}{l}{\wfitwosixlong}
          & A1                    & A2                    & B                     & C                     & \\ \cline{1-6}
$u'$      & $-0.61\pm0.01\pm0.00$ & $-1.30\pm0.01\pm0.00$ & $0\pm0.02$            & $+0.18\pm0.01\pm0.01$ & 0.34 \\
$g'$      & $-0.99\pm0.01\pm0.05$ & $-1.27\pm0.01\pm0.05$ & $0\pm0.02$            & $+0.22\pm0.01\pm0.05$ & 0.20 \\
$r'$      & $-1.25\pm0.01\pm0.05$ & $-1.21\pm0.01\pm0.05$ & $0\pm0.02$            & $+0.22\pm0.01\pm0.05$ & 0.12 \\
$i'$      & $-1.35\pm0.01\pm0.00$ & $-1.17\pm0.01\pm0.00$ & $0\pm0.02$            & $+0.23\pm0.01\pm0.01$ & 0.12 \\
$z'$      & $-1.38\pm0.01\pm0.05$ & $-1.13\pm0.01\pm0.05$ & $0\pm0.02$            & $+0.25\pm0.01\pm0.05$ & 0.12 \\
$J$       & $-1.53\pm0.01\pm0.00$ & $-1.11\pm0.01\pm0.00$ & $0\pm0.02$            & $+0.29\pm0.01\pm0.01$ & 0.18 \\
$H$       & $-1.49\pm0.01\pm0.07$ & $-0.90\pm0.01\pm0.07$ & $0\pm0.02$            & $+0.30\pm0.02\pm0.05$ & 0.19 \\
$K_s$     & $-1.45\pm0.01\pm0.05$ & $-0.83\pm0.02\pm0.05$ & $0\pm0.02$            & $+0.26\pm0.02\pm0.05$ & 0.19 \\
0.5--8keV (ObsID 7758) & $-1.91\pm0.34\pm0.20$ & $-0.75\pm0.71\pm1.55$ & $0\pm0.10$ & $+1.00\pm0.33\pm0.15$ & 0.58 \\
\cline{1-6}\\[-6pt]
\multicolumn{1}{l}{\wfithreelong}
          & A1                    & A2                    & B                     & C                     & \\ \cline{1-6}
$u'$      & $-0.40\pm0.02\pm0.00$ & $+0.07\pm0.03\pm0.00$ & $0\pm0.03$            & $+0.70\pm0.03\pm0.03$ & 0.07 \\
$g'$      & $-0.53\pm0.01\pm0.05$ & $+0.14\pm0.01\pm0.05$ & $0\pm0.03$            & $+0.48\pm0.01\pm0.06$ & 0.06 \\
$r'$      & $-0.52\pm0.01\pm0.00$ & $+0.05\pm0.01\pm0.00$ & $0\pm0.03$            & $+0.38\pm0.01\pm0.03$ & 0.06 \\
$i'$      & $-0.55\pm0.01\pm0.05$ & $-0.05\pm0.01\pm0.05$ & $0\pm0.03$            & $+0.23\pm0.01\pm0.06$ & 0.10 \\
$z'$      & $-0.53\pm0.02\pm0.00$ & $-0.05\pm0.02\pm0.00$ & $0\pm0.03$            & $+0.28\pm0.02\pm0.03$ & 0.08 \\
$J$       & $-0.60\pm0.01\pm0.05$ & $-0.12\pm0.01\pm0.05$ & $0\pm0.03$            & $+0.15\pm0.01\pm0.06$ & 0.12 \\
$H$       & $-0.58\pm0.01\pm0.05$ & $-0.14\pm0.01\pm0.05$ & $0\pm0.03$            & $+0.13\pm0.01\pm0.06$ & 0.13 \\
$K_s$     & $-0.58\pm0.01\pm0.00$ & $-0.10\pm0.02\pm0.00$ & $0\pm0.03$            & $+0.10\pm0.02\pm0.03$ & 0.15 \\
0.5--8keV (ObsID 5603) & $+0.15\pm0.18\pm0.25$ & $+0.00\pm0.22\pm0.35$ & $0\pm0.20$ & $+0.49\pm0.17\pm0.25$ & 0.28 
\enddata
\tablenotetext{a}{HS: highly magnified saddle point; HM: highly
  magnified minimum; LS: less magnified saddle point; LM: less
  magnified minimum.}
\tablenotetext{b}{Standard deviation of the difference of the four image
  magnitudes and the lens model prediction in magnitudes. This
  quantity serves as a quick estimate of the degree of anomaly in the
  flux ratios.}
\tablecomments{All values are in magnitudes. Formal statistical
  uncertainties are reported first, followed by estimated systematic
  uncertainties. Where there are no formal uncertainties, only the
  systematics are reported.}
\end{deluxetable*}

\clearpage
\setcounter{table}{6}
\begin{deluxetable*}{cccccccc}
\tablewidth{0pt}
\tablecaption{Half-light Radii
  \label{tab:sizes}}
\tablehead{
  \colhead{Quasar/Wavelength}&
  \multicolumn{3}{c}{$\log (r_{1/2}/\mathrm{cm})$, Log Prior} & &
  \multicolumn{3}{c}{$\log (r_{1/2}/\mathrm{cm})$, Linear Prior}\\
  \cline{2-4}
  \cline{6-8}
  \colhead{($\mu$m)} &
  \colhead{Median} &
  \colhead{Mode} &
  \colhead{Error} & &
  \colhead{Median} &
  \colhead{Mode} &
  \colhead{Error}
}
\startdata
\multicolumn{1}{l}{\hetwolong} & & & \\ \cline{1-8}
$0.11$ & $16.66$ & $16.78$ & $0.26$ & & $16.75$ & $16.83$ & $0.22$ \\
$0.15$ & $16.80$ & $16.89$ & $0.34$ & & $16.92$ & $16.96$ & $0.29$ \\
$0.20$ & $16.70$ & $16.81$ & $0.34$ & & $16.82$ & $16.87$ & $0.32$ \\
$0.24$ & $16.65$ & $16.75$ & $0.34$ & & $16.77$ & $16.83$ & $0.33$ \\
$0.29$ & $16.43$ & $16.57$ & $0.34$ & & $16.53$ & $16.61$ & $0.30$ \\
$0.40$ & $16.44$ & $16.57$ & $0.36$ & & $16.55$ & $16.62$ & $0.35$ \\
$0.52$ & $16.35$ & $16.35$ & $0.38$ & & $16.47$ & $16.60$ & $0.36$ \\
$0.68$ & $16.29$ & $16.32$ & $0.42$ & & $16.40$ & $16.59$ & $0.40$ \\
\cline{1-8}\\[-6pt]
\multicolumn{1}{l}{\mgfourlong} & & & \\ \cline{1-8}
$0.17$ & $15.38$ & $15.50$ & $0.34$ & & $15.52$ & $15.62$ & $0.31$ \\
$0.21$ & $15.68$ & $15.81$ & $0.45$ & & $15.85$ & $15.95$ & $0.36$ \\
$0.25$ & $15.77$ & $15.92$ & $0.41$ & & $15.94$ & $16.05$ & $0.38$ \\
$0.34$ & $15.96$ & $16.12$ & $0.40$ & & $16.10$ & $16.21$ & $0.34$ \\
$0.45$ & $16.13$ & $16.29$ & $0.43$ & & $16.30$ & $16.40$ & $0.35$ \\
$0.59$ & $16.23$ & $16.36$ & $0.38$ & & $16.37$ & $16.46$ & $0.36$ \\
\cline{1-8}\\[-6pt]
\multicolumn{1}{l}{\hefourlong} & & & \\ \cline{1-8}
$0.13$ & $15.71$ & $15.82$ & $0.42$ & & $15.97$ & $16.03$ & $0.38$ \\
$0.18$ & $15.89$ & $15.98$ & $0.42$ & & $16.15$ & $16.24$ & $0.37$ \\
$0.23$ & $15.93$ & $16.04$ & $0.38$ & & $16.17$ & $16.27$ & $0.32$ \\
$0.29$ & $16.03$ & $16.16$ & $0.37$ & & $16.27$ & $16.37$ & $0.32$ \\
$0.34$ & $16.09$ & $16.26$ & $0.36$ & & $16.31$ & $16.39$ & $0.30$ \\
$0.46$ & $16.17$ & $16.35$ & $0.36$ & & $16.39$ & $16.47$ & $0.31$ \\
$0.61$ & $16.20$ & $16.38$ & $0.36$ & & $16.41$ & $16.49$ & $0.31$ \\
$0.80$ & $16.28$ & $16.43$ & $0.34$ & & $16.47$ & $16.52$ & $0.33$ \\
\cline{1-8}\\[-6pt]
\multicolumn{1}{l}{\rxninelong} & & & \\ \cline{1-8}
$0.09$ & $16.14$ & $16.20$ & $0.32$ & & $16.30$ & $16.29$ & $0.33$ \\
$0.13$ & $16.10$ & $16.17$ & $0.31$ & & $16.25$ & $16.26$ & $0.31$ \\
$0.17$ & $16.13$ & $16.19$ & $0.32$ & & $16.29$ & $16.27$ & $0.34$ \\
$0.20$ & $16.16$ & $16.21$ & $0.30$ & & $16.32$ & $16.31$ & $0.34$ \\
$0.24$ & $16.19$ & $16.24$ & $0.30$ & & $16.35$ & $16.34$ & $0.33$ \\
$0.33$ & $16.26$ & $16.32$ & $0.29$ & & $16.40$ & $16.40$ & $0.32$ \\
$0.43$ & $16.17$ & $16.25$ & $0.30$ & & $16.31$ & $16.33$ & $0.32$ \\
$0.57$ & $16.25$ & $16.32$ & $0.27$ & & $16.37$ & $16.38$ & $0.28$ \\
\cline{1-8}\\[-6pt]
\multicolumn{1}{l}{\sdssninelong} & & & \\ \cline{1-8}
$0.14$ & $15.78$ & $15.92$ & $0.35$ & & $15.93$ & $16.05$ & $0.31$ \\
$0.19$ & $15.65$ & $15.80$ & $0.33$ & & $15.78$ & $15.91$ & $0.28$ \\
$0.25$ & $15.66$ & $15.83$ & $0.32$ & & $15.78$ & $15.92$ & $0.26$ \\
$0.31$ & $15.68$ & $15.85$ & $0.32$ & & $15.79$ & $15.93$ & $0.26$ \\
$0.36$ & $15.68$ & $15.84$ & $0.32$ & & $15.80$ & $15.93$ & $0.26$ \\
$0.50$ & $15.75$ & $15.87$ & $0.37$ & & $15.90$ & $15.98$ & $0.31$ \\
$0.65$ & $15.79$ & $15.91$ & $0.39$ & & $15.94$ & $16.02$ & $0.32$ \\
$0.86$ & $15.84$ & $15.97$ & $0.34$ & & $16.00$ & $16.08$ & $0.30$ \\
\cline{1-8}\\[-6pt]
\multicolumn{1}{l}{\heoneonelong} & & & \\ \cline{1-8}
$0.21$ & $15.78$ & $15.96$ & $0.31$ & & $15.93$ & $16.04$ & $0.23$ \\
$0.28$ & $15.84$ & $16.02$ & $0.33$ & & $15.99$ & $16.11$ & $0.23$ \\
$0.34$ & $15.84$ & $16.02$ & $0.33$ & & $15.99$ & $16.10$ & $0.23$ \\
$0.41$ & $15.78$ & $15.94$ & $0.30$ & & $15.91$ & $16.03$ & $0.23$ \\
$0.56$ & $15.75$ & $15.93$ & $0.31$ & & $15.90$ & $16.01$ & $0.23$ \\
$0.74$ & $15.83$ & $16.01$ & $0.31$ & & $15.97$ & $16.09$ & $0.22$ \\
$0.97$ & $15.87$ & $16.05$ & $0.34$ & & $16.02$ & $16.14$ & $0.23$ \\
\cline{1-8}\\[-6pt]
\multicolumn{1}{l}{\pgoneonelong} & & & \\ \cline{1-8}
$0.13$ & $16.34$ & $16.49$ & $0.32$ & & $16.48$ & $16.57$ & $0.25$ \\
$0.18$ & $16.65$ & $16.72$ & $0.37$ & & $16.78$ & $16.81$ & $0.35$ \\
$0.23$ & $16.69$ & $16.79$ & $0.30$ & & $16.81$ & $16.85$ & $0.28$ \\
$0.28$ & $16.76$ & $16.82$ & $0.35$ & & $16.89$ & $16.89$ & $0.32$ \\
$0.33$ & $16.74$ & $16.82$ & $0.30$ & & $16.86$ & $16.89$ & $0.29$ \\
$0.46$ & $16.74$ & $16.81$ & $0.34$ & & $16.86$ & $16.86$ & $0.31$ \\
$0.61$ & $16.63$ & $16.73$ & $0.32$ & & $16.75$ & $16.79$ & $0.27$ \\
$0.79$ & $16.84$ & $16.88$ & $0.32$ & & $16.96$ & $17.11$ & $0.29$ \\
\pagebreak
\cline{1-8}\\[-6pt]
\multicolumn{1}{l}{\rxoneonelong} & & & \\ \cline{1-8}
$0.22$ & $15.35$ & $15.45$ & $0.35$ & & $15.62$ & $15.73$ & $0.40$ \\
$0.29$ & $15.41$ & $15.54$ & $0.47$ & & $15.69$ & $15.81$ & $0.35$ \\
$0.38$ & $15.36$ & $15.49$ & $0.47$ & & $15.66$ & $15.78$ & $0.35$ \\
$0.46$ & $15.48$ & $15.62$ & $0.49$ & & $15.75$ & $15.86$ & $0.35$ \\
$0.55$ & $15.57$ & $15.70$ & $0.42$ & & $15.81$ & $15.90$ & $0.31$ \\
$0.75$ & $15.59$ & $15.69$ & $0.32$ & & $15.77$ & $15.87$ & $0.26$ \\
$1.00$ & $15.58$ & $15.67$ & $0.29$ & & $15.75$ & $15.84$ & $0.25$ \\
$1.30$ & $15.66$ & $15.77$ & $0.32$ & & $15.85$ & $15.94$ & $0.28$ \\
\cline{1-8}\\[-6pt]
\multicolumn{1}{l}{\sdssoneonelong} & & & \\ \cline{1-8}
$0.10$ & $15.78$ & $15.88$ & $0.33$ & & $15.95$ & $16.06$ & $0.30$ \\
$0.14$ & $15.92$ & $16.04$ & $0.43$ & & $16.13$ & $16.19$ & $0.34$ \\
$0.18$ & $15.92$ & $16.04$ & $0.39$ & & $16.12$ & $16.19$ & $0.31$ \\
$0.22$ & $15.92$ & $16.04$ & $0.37$ & & $16.12$ & $16.20$ & $0.31$ \\
$0.26$ & $15.96$ & $16.09$ & $0.51$ & & $16.17$ & $16.24$ & $0.40$ \\
$0.36$ & $15.96$ & $16.10$ & $0.38$ & & $16.17$ & $16.25$ & $0.31$ \\
$0.48$ & $16.01$ & $16.16$ & $0.40$ & & $16.22$ & $16.31$ & $0.33$ \\
$0.63$ & $16.23$ & $16.35$ & $0.45$ & & $16.43$ & $16.47$ & $0.37$ \\
\cline{1-8}\\[-6pt]
\multicolumn{1}{l}{\sdssonethreelong\tnm{a}} & & & \\ \cline{1-8}
$0.15$ & $15.96$ & \nodata & \nodata & &$16.22$ & \nodata & \nodata \\
$0.20$ & $15.98$ & \nodata & \nodata & &$16.25$ & \nodata & \nodata \\
$0.26$ & $16.15$ & \nodata & \nodata & &$16.42$ & \nodata & \nodata \\
$0.32$ & $16.11$ & \nodata & \nodata & &$16.38$ & \nodata & \nodata \\
$0.38$ & $16.05$ & \nodata & \nodata & &$16.33$ & \nodata & \nodata \\
$0.52$ & $16.24$ & \nodata & \nodata & &$16.49$ & \nodata & \nodata \\
$0.69$ & $16.23$ & \nodata & \nodata & &$16.49$ & \nodata & \nodata \\
$0.90$ & $16.23$ & \nodata & \nodata & &$16.55$ & \nodata & \nodata \\
\cline{1-8}\\[-6pt]
\multicolumn{1}{l}{\wfitwosixlong} & & & \\ \cline{1-8}
$0.11$ & $15.73$ & $15.83$ & $0.52$ & & $16.07$ & $16.08$ & $0.44$ \\
$0.15$ & $16.20$ & $16.31$ & $0.44$ & & $16.46$ & $16.56$ & $0.37$ \\
$0.20$ & $16.47$ & $16.56$ & $0.32$ & & $16.61$ & $16.67$ & $0.27$ \\
$0.24$ & $16.79$ & $16.88$ & $0.34$ & & $16.91$ & $16.98$ & $0.29$ \\
$0.28$ & $16.52$ & $16.63$ & $0.30$ & & $16.65$ & $16.73$ & $0.25$ \\
$0.39$ & $16.36$ & $16.53$ & $0.32$ & & $16.51$ & $16.61$ & $0.22$ \\
$0.51$ & $16.34$ & $16.52$ & $0.47$ & & $16.56$ & $16.69$ & $0.29$ \\
$0.67$ & $16.27$ & $16.44$ & $0.47$ & & $16.53$ & $16.67$ & $0.30$ \\
\cline{1-8}\\[-6pt]
\multicolumn{1}{l}{\wfithreelong} & & & \\ \cline{1-8}
$0.14$ & $16.86$ & $16.98$ & $0.36$ & & $16.97$ & $17.04$ & $0.28$ \\
$0.18$ & $16.89$ & $16.96$ & $0.40$ & & $17.02$ & $17.04$ & $0.35$ \\
$0.24$ & $16.84$ & $16.91$ & $0.37$ & & $16.96$ & $17.01$ & $0.31$ \\
$0.29$ & $16.58$ & $16.63$ & $0.38$ & & $16.71$ & $16.81$ & $0.31$ \\
$0.34$ & $16.65$ & $16.69$ & $0.38$ & & $16.78$ & $16.87$ & $0.30$ \\
$0.47$ & $16.51$ & $16.58$ & $0.39$ & & $16.66$ & $16.75$ & $0.30$ \\
$0.62$ & $16.47$ & $16.54$ & $0.37$ & & $16.61$ & $16.70$ & $0.29$ \\
$0.81$ & $16.44$ & $16.52$ & $0.36$ & & $16.58$ & $16.69$ & $0.28$
\enddata
\tablenotetext{a}{90\% upper limits on the half-light radius for
\sdssonethree. Because our priors cut off below a certain source size,
these limits are conservative.}
\end{deluxetable*}


\typeout{get arXiv to do 4 passes: Label(s) may have changed. Rerun}

\end{document}